\documentclass{jfm}
\usepackage{appendix}

\usepackage{graphicx}
\usepackage{newtxtext}
\usepackage{bm}
\usepackage{newtxmath}
\usepackage{natbib}
\usepackage{hyperref}
\usepackage{graphicx}
\hypersetup{
    colorlinks = true,
    urlcolor   = blue,
    citecolor  = black,
}

\newcommand{\RomanNumeralCaps}[1]

\graphicspath{{figures/}}

\title{Experimental Investigation of Finite-size Spheres Settling in Homogeneous Turbulence}

\author{Joo Young Bang\aff{1}
  \corresp{\email{jbang9@wisc.edu}}
  \and Gautier Verhille\aff{2}
  \and Nimish Pujara\aff{1,3} 
  }
  
\affiliation{\aff{1} Department of Civil and Environmental Engineering, University of Wisconsin-Madison, Madison, WI 53706, USA
\aff{2} Aix-Marseille Université, CNRS, Centrale Méditerranée, IRPHE, Marseille, 13013, France
\aff{3}Department of Civil Engineering, The University of British Columbia, Vancouver, BC V6T 1Z4, Canada}

\begin{document}
\maketitle

\begin{abstract}
We investigate the dynamics of finite-size spheres settling in homogeneous turbulence using simultaneous measurements of particle motion and fluid flow over a measurement volume larger than the integral scale. For weakly settling particles, particle acceleration statistics are similar to previous results for neutrally buoyant finite-size particles. For strongly settling particles, particle acceleration statistics show a strong imprint of particle size and terminal settling velocity. In particular, particles filter out turbulent fluid motions below their size and encounter new turbulent motions due to settling via the crossing-trajectory effect. Apart from settling, particle motion is strongly correlated with the local flow coarse-grained at the integral scale, with this behaviour explained by scaling analysis that shows strongly settling particles always have sufficient time to relax to the integral-scale fluid motions. We also find that the particle settling velocity in turbulence can be modified from the terminal settling velocity in quiescent fluid. A new framework to understand this settling modulation accounts for large-scale background flow and decomposes net settling modulation into changes in mean particle-fluid relative velocity and preferential sampling. While we do not observe significant preferential sampling, reduced settling with respect to the local flow is best predicted by different regimes of the particle Reynolds number based on the terminal settling velocity. Data from particles with Reynolds numbers above the path-instability threshold suggest that turbulence-induced modifications of non-axisymmetric wakes may increase the upward hydrodynamic force, ensuring particles to settle at lower velocities relative to the local fluid motion.
\end{abstract}

\section{Introduction}
\label{sec:Introduction}

Particle settling in turbulence is important as the collective particle flux governs a wide range of large-scale environmental processes. For example, turbulence in the upper ocean influences the aggregation and fragmentation of marine organic particles, thereby modifying their size and settling rates and ultimately affecting the vertical transport of carbon from the surface to the deep ocean \citep{Burd_2009,Takeuchi_2019}. Similarly, the settling of droplets and ice crystals in atmospheric turbulence plays a central role in cloud formation and precipitation \citep{Pumir_2016}. 

In natural aquatic environments, settling particles typically have densities only marginally greater than the density of the surrounding water. Examples include zooplankton fecal pellets and negatively buoyant plastics, which commonly have specific gravities of approximately $1.04 - 2.30$ \citep{Atkinson_2012, Cole_2016, Poulain-Zarcos24}, and quartz-rich sediments with a specific gravity of approximately 2.65 \citep{garcia2008sedimentation}. For particles within this density range, the relative importance of settling and flow-induced particle motion is very sensitive to particle size. Sufficiently small particles will remain in suspension and follow the most energetic fluid motions, whereas the motion of larger particles is likely to be settling-dominated. Between these limits, both settling and the surrounding flow influence particle motion. This intermediate regime is particularly relevant to millimetre-scale particles in aquatic environments because their settling velocities in quiescent water can be comparable in magnitude to flow velocities encountered in rivers and the surface mixed layer in lakes and oceans \citep{Sutherland_2015,D_jeans_2022,DCP2022enhanced}. For typical turbulent energy dissipation rates, these particles will also have diameters that exceed the Kolmogorov length scale of turbulence \citep{Sutherland_2015, D_jeans_2022}, so that their size influences how they interact with turbulent motions at different scales. In addition, their particle Reynolds numbers (defined using the terminal settling velocity and particle size) are typically greater than one and can reach up to $10^3$ \citep{DCP2022enhanced}, showing that they can span different drag regimes. Thus, millimetre-scale particles in natural aquatic environments are inertial, finite-size settling particles, where the inertia due to the fluid and the particle are both important. 

The dynamics of inertial, finite-size particles settling in turbulent flows are not yet fully understood, even for spherical particles in dilute suspensions where particle-particle interactions and collective feedback to the turbulence are negligible. Several questions are particularly relevant to the present study. First, turbulence is known to modify the net settling velocity of finite-size particles relative to that in quiescent fluid \citep[\textit{e.g.},][]{Fornari_2016_reduced, akutina2020experimental}, but it remains unclear which particle and flow properties govern the direction and magnitude of this modification and what physical mechanisms underlie it. Second, the effects of particle size on its acceleration statistics for neutrally buoyant finite-size particles are reasonably well established \citep[\textit{e.g.},][]{voth2002measurement, Qureshi_2007, VOLK_2011}, but how settling modifies these statistics remains unclear. Understanding particle acceleration is important because it is directly proportional to the net force acting on the particles. More broadly, both questions are related to the issue of understanding which scales of turbulence govern the particle dynamics. 

Addressing these questions remains challenging for several reasons. Direct comparisons across studies are hampered by different choices of which dimensionless parameters to vary and over what range. Further limitations arise from the fact that, unlike sub-Kolmogorov particles, whose motion can be described using point-particle models such as the Maxey--Riley--Gatignol equation, no broadly applicable model exists for finite-size particles in turbulence. Numerical studies have employed immersed-boundary methods to resolve fluid-particle interactions directly \citep{Fornari_2016_Sedimentation, Fornari_2016_reduced, Fornari_2018_settling, Chiarini_2024}, but these particle-resolved simulations remain computationally demanding, limiting the range of particle and flow conditions that can be systematically investigated \citep{Brandt_2022}. In laboratory experiments, the challenge is not being able to measure all variables of interest. It is common to either measure snapshots of individual particles and the surrounding flow or to measure particle trajectories without simultaneous measurements of the surrounding flow. In both cases, there is limited information on how particle motion co-evolves with the surrounding flow.

Here, we use laboratory experiments to study how finite-size, marginally denser than water spherical particles settle in homogeneous turbulence where we simultaneously measure both particle motion and the fluid motion over a measurement volume larger than the turbulent integral scale. We find that turbulence reduces the settling velocity of particles above the Reynolds number threshold associated with the onset of path instability in quiescent fluid, where this reduction manifests as a reduction in the settling rate relative to the particle\textquotesingle s local flow. We also find that the particle acceleration is affected by both finite-size and settling, with particles filtering out eddies smaller than their size but experiencing more frequent encounters with turbulent eddies due to settling. Together, our results suggest that finite-size and settling combine to determine how particles interact with turbulent eddies and how these interactions govern their motion.

In the remainder of the paper, \S\ref{sec:Background} provides further background information, \S\ref{sec:Experimental_setup} describes the experimental methods, \S\ref{sec:Results} presents the results, and finally, \S\ref{sec:Conclusions} summarises the main conclusions.

\section{Background}
\label{sec:Background}

\subsection{Dimensionless parameters}
\label{sec:Background_Dimensionless_Parameters}
The settling of finite-size, marginally denser than the fluid spherical particles in homogeneous turbulence is governed by the following dimensional variables. The particle is characterised by its diameter $d_p$, density $\rho_p$, and terminal settling velocity in quiescent fluid $v_s$, while the turbulence is characterised by its (large-scale) velocity fluctuation scale $u_L$ and mean turbulent kinetic energy dissipation rate $\varepsilon$. Together with the fluid density $\rho$, kinematic viscosity $\nu$, and gravitational acceleration $\mathit{g}$, there are eight dimensional variables with three fundamental dimensions, yielding five independent dimensionless parameters. We include the particle terminal settling velocity $v_s$ explicitly since it cannot be predicted \textit{a priori} for finite particle Reynolds number \citep[see the analogous discussion in][which also considers the Galileo number as an alternative measure of fluid inertia]{Bang_2025}. In addition, other properties of the turbulent flow such as the integral length scale $L$, Taylor microscale $\lambda$, and Kolmogorov length, time, and velocity scales ($\eta$, $\tau_\eta$, and $u_\eta$, respectively) can be determined from the above variables assuming isotropy \citep[Ch.~6.3][]{Pope_2000}. 

We define the five independent dimensionless parameters following the same approach as in \citet{Bang_2025}: we split the dimensionless numbers as those for characterising (a) particle settling in the absence of any flow, (b) the characteristics of the turbulent flow without the particles, and (c) the interaction of the particles with the flow. The particle-to-fluid density ratio, $\gamma = \rho_p/\rho$, and terminal settling velocity particle Reynolds number, $Re_{p,t} = d_p v_s/\nu$ characterise particle settling in quiescent fluid \citep{Horowitz2010effect, Raaghav2022path, Cabrerebooman2024path}. Then, the scale separation of the turbulent flow is characterised by Taylor microscale Reynolds number $Re_\lambda = u_L \lambda /\nu = u_L^2 \sqrt{30/\nu\varepsilon}$ \citep{Pope_2000}. The particle interactions with the turbulence are characterised by the strength of the Kolmogorov-scale flow accelerations ($a_\eta = \varepsilon^{3/4}/\nu^{1/4}$) relative to gravity via the Froude number, $Fr_\eta = \varepsilon^{3/4}/\mathit{g} \nu^{1/4}$, and the relative importance of particle settling to the turbulent fluctuations of velocity via the settling number, $Sv_L = v_s/u_L$. Together, the five dimensionless numbers are  
\begin{equation}
\gamma = \frac{\rho_p}{\rho}; \,\,\, Re_{p,t} = \frac{d_p v_s}{\nu}; \,\,\, Re_\lambda = u_L^2 \sqrt{\frac{30}{\nu\varepsilon}}; \,\,\, Fr_\eta = \frac{\varepsilon^{3/4}}{\mathit{g}\nu^{1/4}}; \,\,\, Sv_L = \frac{v_s}{u_L},
\label{eq:dimensionless_params}
\end{equation}
where the subscript $\eta$ denotes use of a Kolmogorov-scale turbulence quantity whereas the subscript $L$ indicates use of an integral-scale turbulence quantity.

Other widely used dimensionless parameters can be expressed using these independent parameters. The size number, which represents the particle size relative to the Kolmogorov length scale, is defined as
\begin{equation}
    Sz_\eta = \frac{d_p}{\eta} = \frac {30^{1/4}Re_{p,t}}{Sv_L Re_\lambda^{1/2}}.
    \label{eq:Size_number}
\end{equation}

The Stokes number, which characterises particle inertia, is usually formulated as the ratio of a particle timescale to the Kolmogorov timescale. We consider two candidate particle timescales. The first is the response time based on the Schiller--Naumann drag correlation for a sphere, $\tau_{p,SN} = d_p^2 (\gamma+0.5)/18\nu(1+0.15 Re_{p,t}^{0.687})$, which accounts for finite-particle-Reynolds-number corrections to Stokes drag assuming quasi-steady flow around the particle. This timescale approximates the characteristic relaxation time for a particle to approach its terminal settling velocity when the history force can be neglected \citep[see][for recent experimental evidence of this]{Bang_2025}. The second candidate particle timescale is the settling-based timescale, $d_p/v_s$, which represents the time required for a particle to settle a distance equal to its own size at its terminal settling velocity. This timescale characterises the eddy crossing time for turbulent eddies of the same size as the particle due to settling. The corresponding Stokes numbers based on these two particle timescales are
\begin{align}
    St_{\eta,SN} &= \frac{\tau_{p,SN}}{\tau_\eta} = \frac{30^{1/2}(\gamma+0.5) Re_{p,t}^2}{18(1+0.15 Re_{p,t}^{0.687}) Sv_L^2 Re_\lambda}, 
    \label{eq:Stokes_number_SN} \\
    St_{\eta,d_p/v_s} &= \frac{d_p/v_s}{ \tau_\eta} = \frac{Sz_\eta^2}{Re_{p,t}} = \frac{30^{1/2} Re_{p,t}}{Sv_L^2 Re_\lambda}.
    \label{eq:Stokes_number_dpvs}
\end{align}

The Froude number can be modified to better reflect the influence of gravity on the particle by replacing the gravitational acceleration with the gravitational acceleration of the submerged weight, $(\gamma-1)\textrm{g}/\gamma$ (i.e. the buoyancy-corrected weight per unit particle mass). We write this as the `particle submerged weight Froude number'
\begin{equation}
    Fr_{\eta,p} = \frac{\gamma \, Fr_{\eta}}{\left(\gamma - 1 \right)} = \frac{\gamma\varepsilon^{3/4}}{\left(\gamma - 1 \right)\mathit{g}\nu^{1/4}}.
    \label{eq:Froude_number_particle}
\end{equation}

\subsection{Particle acceleration}
\label{sec:Background_Particle_Acceleration}

Particle size is known to affect its acceleration statistics in turbulence. In laboratory experiments of finite-size spheres with negligible settling ($Sv_L \ll1$), the particle acceleration variance was found to scale as $d_p^{-2/3}$ up to $Sz_\eta = 45$. This was attributed to the inertial-range scaling of pressure increments at the scale of the particle diameter based on the Kolmogorov-Richardson phenomenology, with a hint of an intermittency correction \citep{voth2002measurement,Qureshi_2007,VOLK_2011}. The effect of particle size was also evident in the particle response time (estimated from the decay of the acceleration autocorrelation) and in acceleration intermittency (quantified by the acceleration kurtosis). For increasing particle size, the response time increased and the acceleration kurtosis decreased, with these findings rationalised as being caused by how particles filter the surrounding turbulence at the scale of their diameter, with larger particles being insensitive to smaller-scale turbulent fluctuations \citep{Voth_1998,voth2002measurement,VOLK_2011}.

However, it has also been found that acceleration statistics can be substantially different for settling particles. \citet{Bourgoin_2011} examined finite-size particles with $12 \leq Sz_\eta \leq 25$ and $1 \le \gamma \le 65$ settling in a wind tunnel and found that their acceleration variance could not be explained by particle size alone and that the particle response times remained nearly constant across different particle sizes and densities. \citet{Ireland_2016} and \citet{Berk_2021} examined sub-Kolmogorov particles in simulations and experiments, respectively, and found that settling-induced modifications to acceleration statistics could be explained through the crossing-trajectories effect, whereby settling causes particles to cross turbulent eddies more rapidly and sample faster variations in the fluid velocity \citep{Yudine59}. This effect can enhance particle acceleration fluctuations, although it was also found that particle inertia may counteract this enhancement. It is not known whether finite-size settling particles would experience a similar effect, and if they do, which turbulent scales would be responsible.

To better understand the role of settling on particle acceleration statistics, we write the equation of motion for finite-size settling particles in turbulence as \citep[cf.][]{Fornari_2016_reduced} 
\begin{equation}
        m_p\bm{a} = -(m_p-m_f)\textrm{g}\bm{e}_z + \bm{F}_{\textrm{hd}}, \label{eq:particle_force}
\end{equation}
where $\bm{a}$ is the particle acceleration vector, $m_p$ is the particle mass, $m_f$ is the fluid mass displaced by the particle, and $\bm{e}_z$ is the unit vector pointing upwards against gravity. The force on the particle is the sum of its weight and the surface integral of the pressure and viscous stresses. We have decomposed the pressure contribution into the upward `hydrostatic' buoyancy force $m_f \mathrm{g}$ (also known as the Archimedes force) and an unknown `hydrodynamic' component. The surface integrated viscous and hydrodynamic pressure stresses is then written as $\bm{F}_\textrm{hd}$, the `hydrodynamic force vector' acting on the particle. This hydrodynamic force does not have an analytical expression in terms of the undisturbed local flow for finite-sized settling particles \citep[cf.][who provide a low-Reynolds-number expansion for the force for small particles]{MaxeyRiley1983}. Dividing throughout by the particle mass gives the particle acceleration to be
\begin{equation}
        \bm{a} = -\left(\frac{\gamma - 1}{\gamma}\right)\textrm{g} \bm{e}_z + \bm{a}_{\textrm{hd}}, \label{eq:particle_acceleration}
\end{equation}
where $\bm{a}_\textrm{hd}=\bm{F}_\textrm{hd}/m_p$. The first term in Eq.~\eqref{eq:particle_acceleration} then provides a useful baseline acceleration scale for particle motion, which is why it is used to make the turbulence acceleration scale dimensionless in $Fr_{\eta,p}$ in Eq.~\eqref{eq:Froude_number_particle}.

\subsection{Net settling modulation}
\label{sec:Background_Net_settling_modulation}

Turbulence is known to modify the net settling velocity of particles relative to that in quiescent fluid, with several mechanisms proposed to explain this modification. Settling enhancement has been attributed to preferential sweeping, whereby particles preferentially sample downward-moving flow regions outside vortex cores \citep{Nielsen_1993}. In contrast, reduced settling has been associated with (a) vortex trapping, in which small, weakly settling particles become trapped within turbulent vortical structures \citep{Tooby_1977, Nielsen_1984,Nielsen_1993}; (b) loitering, in which particles spend more time in upward-flow regions than in downward-flow regions \citep{Nielsen_1993}; and (c) nonlinear drag, whereby fluctuations in the particle-to-fluid relative velocity increase the mean drag force \citep{Nielsen_1993, Wang_Maxey_1993}. 

Settling modulation has been studied extensively for sub-Kolmogorov and Kolmogorov-scale particles ($Sz_\eta \lesssim1$) in laboratory experiments \citep{Nielsen_1993, ALISEDA_2002, Good_2014, Mora2021, Ferran2023, Clementi_2026} and numerical simulations \citep{Wang_Maxey_1993, Good_2014, Ireland_2016,  Keane_2025}. Settling enhancement and reduction have both been reported, with the enhancement linked to preferential sweeping and the reduction to loitering and nonlinear drag, and a transition between the two observed as $Sv_L$ increases \citep{Good_2014,Mora2021,Ferran2023,Clementi_2026}. The $Sv_L$ value at which this transition occurs has been reported to vary with $Re_\lambda$ \citep{Mora2021,Ferran2023}.

In contrast, settling of finite-size particles ($Sz_\eta \gtrsim 1$) has received less attention. \citet{Fornari_2016_Sedimentation} numerically investigated finite-size spheres ($Sz_\eta = 11.9$, $\gamma = 1.02$) in homogeneous isotropic turbulence ($Re_\lambda =90$) and observed reduced settling. \citet{Fornari_2016_reduced} subsequently extended the analysis to spherical particles of the same size but with varying density ratios ($1<\gamma <1.04$, corresponding to $0.19 < Sv_L < 4.81$) under the same turbulence conditions at dilute particle concentrations, also finding an increasing reduction in particle settling with decreasing $\gamma$ (and $Sv_L$). They analysed this settling reduction by modelling the hydrodynamic force $\bm{F}_\textrm{hd}$ in Eq.~\eqref{eq:particle_force} as the sum of a `steady' drag written as a function of the instantaneous relative velocity of the particle with respect to the local fluid and an `unsteady' drag that serves to capture any remaining force inferred from particle acceleration data. Under this analysis framework, they attributed the reduced particle settling to an increase in the `steady' component of the drag force caused primarily by an increasing intensity in the fluctuations of particle velocity, including the fluctuations relative to the local fluid velocity. This contradiction of unsteady particle motion contributing to the `steady' component of the drag was rationalised by using the instantaneous relative velocity in the Schiller--Naumann drag correlation under the assumption of quasi-steadiness, but the validity of this assumption was not evaluated. 

\citet{akutina2020experimental} experimentally investigated the settling of a finite-size sphere ($\gamma = 1.2$, $4 < Sz_\eta < 20$, and $4 < Sv_L < 65$) in a column of weak, homogeneous, nearly isotropic turbulence generated by two facing oscillating grids. Like \citet{Fornari_2016_reduced}, they also found that the reduction in settling velocity became more pronounced as $Sv_L$ decreased. Given that previous results for sub-Kolmogorov particles ($Sz_\eta \lesssim1$) suggest that the settling reduction should become weaker with decreasing $Sv_L$, these trends for finite-size particles ($Sz_\eta \gtrsim 1$) suggest that finite-size effects may change the phenomenology of settling modulation of particles in turbulence and that a more systematic investigation is necessary. Additionally, it is not clear how important each of the four mechanisms given above is for the settling modulation of finite-size particles across the $Sz_\eta$--$Sv_L$ parameter space.

To add to the complexity, different studies compute different particle velocity statistics to quantify the modified particle settling in turbulence. If all data from different particle trajectories are pooled together, greater weight is given to slow-moving particles with longer trajectories, creating a bias. This problem can be avoided by using small measurement volumes \citep{ALISEDA_2002, Mora2021, Ferran2023} or by using low sampling frequencies in large measurement volumes \citep{Clementi_2026}, so that one obtains approximately one velocity measurement per trajectory. The disadvantage of this approach is that it does not allow access to temporal correlations. Alternatively, one can take the mean or median of each trajectory before pooling data across trajectories \citep{Good_2014, pujara2021measurements}, but this approach does not make full use of all the available data. 

Another important issue is the large-scale background flow, with spatial and temporal scales comparable to or larger than the measured particle trajectories. Examples include both steady \citep{Blum10,Masuk_2019} and unsteady \citep{variano2008random, pujara2021measurements, Bang_2025} tank-scale flows. If only particle velocities are measured, the unknown background vertical flow that may be different for each particle trajectory may lead to a misleading estimate of the particle settling modulation. The background flow may affect the results even in facilities designed to produce negligible mean flows since each particle trajectory experiences a different background flow and the measured particles may be biased toward certain flow patterns. Additionally, the mean flow magnitude may be comparable to the settling modulation. While some studies attempt to correct for this by making flow measurements in particle-free experiments \citep{akutina2020experimental, Ferran2023}, the separately measured or long-time-averaged fluid velocities may not accurately represent the background flow for individual particle trajectories.

To account for these considerations, we propose quantifying the net settling modulation, $\Delta_{NS}$, as the difference between the particle settling velocity in turbulence after correcting for the background flow and the terminal settling velocity in quiescent fluid:
\begin{equation}
    \Delta_{NS} \equiv \langle v_z - \overline{u}_z +v_s\rangle = \langle v_z -  \tilde{u}_z +v_s \rangle + \langle \tilde{u}_z - \overline{u}_z \rangle.  \label{eq:net_settling_quantification}
\end{equation}
Here, ${\bm{v}}$ is the particle velocity vector, ${\bm{u}}({\bm{x}},t)$ is the fluid velocity vector field, and the subscript $z$ denotes the vertical component. Then, $\overline{\bm{u}}$ and $\tilde{\bm{u}}$ represent the large-scale background flow and the local turbulent flow experienced by the particle, respectively, computed by filtering the fluid velocity field at a scale of the measurement volume and at the turbulence integral scale, respectively. By filtering at the integral scale centred around the particle to compute $\tilde{\bm{u}}$, we capture the most energetic flow of turbulent eddies sampled by the particle. And by filtering at the scale of the measurement volume to compute $\overline{\bm{u}}$, we capture the background flow in which each particle trajectory was measured. The $\langle \cdot \rangle$ denotes an ensemble average over particle trajectories, with data from each trajectory contributing equally to the ensemble statistics. Positive and negative $\Delta_{NS}$ values indicate reduced and enhanced settling, respectively.

We also propose a kinematic decomposition of the net settling modulation on the right-hand side of Eq.~\eqref{eq:net_settling_quantification}. The first term quantifies the change in the mean particle-fluid relative velocity, reflecting how the particle-turbulence interactions affect the mean settling relative to the local flow. This change is expected to be related to the hydrodynamic force acting on the particle, as discussed in \S\ref{sec:Background_Particle_Acceleration}. The second term quantifies the extent to which particles preferentially sample upward- or downward-moving fluid velocities in turbulent eddies (\textit{i.e.}, loitering or fast tracking, respectively) after correcting for any large-scale background flow. This framework provides a quantitative basis for evaluating the relative roles of different mechanisms that result in net settling modulation for finite-sized particles in turbulence.

\section{Experimental setup}
\label{sec:Experimental_setup}

\subsection{Turbulence tank and flow characterisation}
\label{sec:turbulence_tank_and_flow_characterisation}
\begin{figure}
    \centering
    \includegraphics[width=1\linewidth]{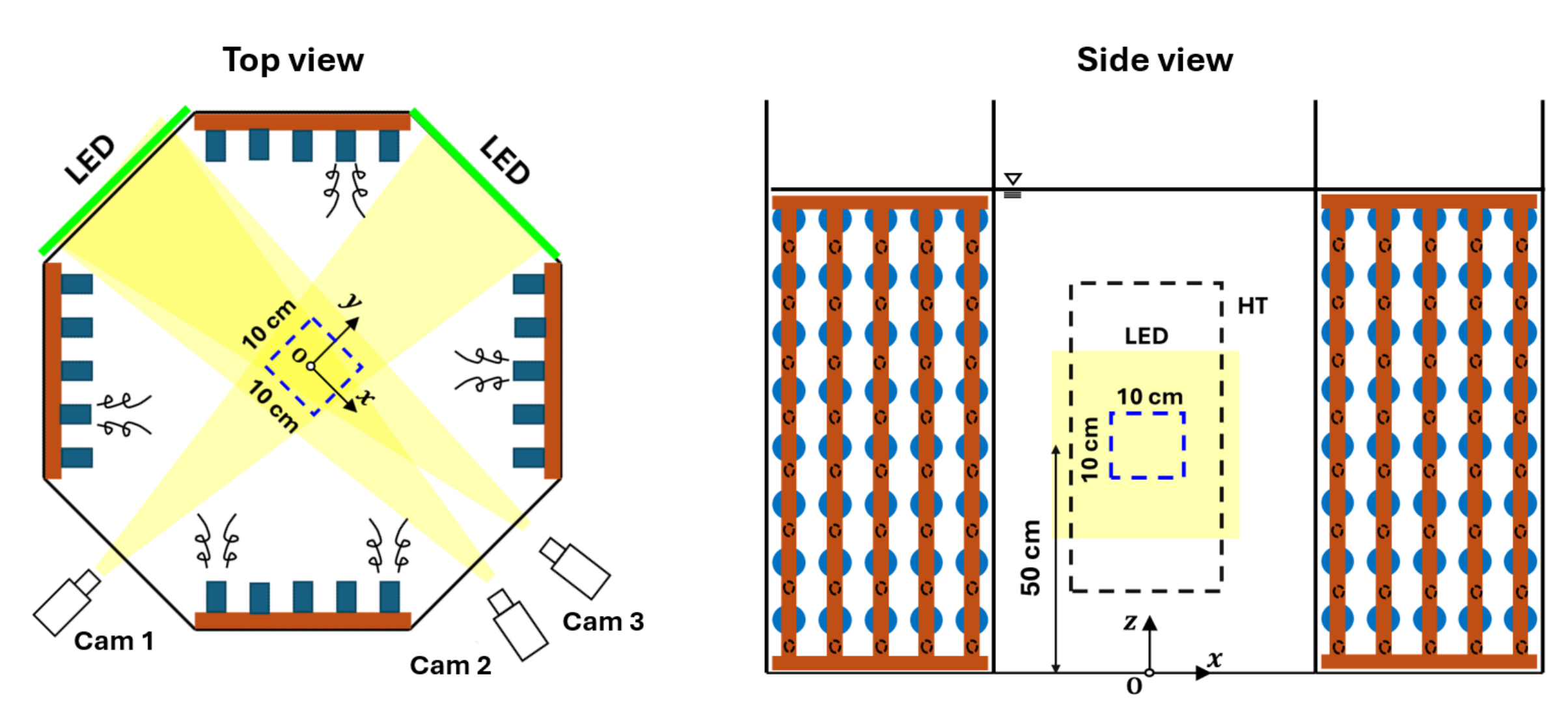}
    \caption{Octagonal tank with four jet-arrays. Homogeneous turbulence (HT) region and measurement volume are illustrated by the black and blue dashed boxes, respectively.}
    \label{fig:ext_setup}
\end{figure}

We conducted laboratory experiments in a vertical octagonal prism tank with a side length of 0.61 m and a height of 1.22 m, filled to a water depth of 1 m (figure \ref{fig:ext_setup}). During the experiments, the water temperature remained at approximately $18^\circ \mathrm{C}$, corresponding to a kinematic viscosity of $\nu = 1.05 \times 10^{-6}$ m$^2$ s$^{-1}$. The origin of the laboratory coordinate system is located at the centre of the tank bottom, with the $x$- and $y$-axes defining the horizontal plane and the $z$-axis directed vertically upward, opposite to gravity. 

Four randomly controlled jet arrays, each consisting of a 5 $\times$ 8 grid of submerged bilge pumps equipped with mesh-grid flow straighteners, generate homogeneous turbulence at the tank centre \citep[see][for a detailed description and characterisation of the turbulent flow in this facility]{Bang_2023}. Briefly, each jet is controlled using the sunbathing algorithm introduced by \citet{variano2008random}, in which each pump is turned on and off independently with durations sampled from a Gaussian distribution whose standard deviation is equal to one-third of its mean. As shown in \citet{Bang_2023}, signatures of individual jets can be detected in the velocity spectra at the tank centre when the dimensionless mean on-time is greater than 4, where the dimensionless mean on-time is the mean on-time multiplied by the jet exit velocity divided by the horizontal distance from the jets to the tank centre. In this study, we chose jet forcing parameters to minimise these jet signatures as these would cause particles to be swept out of the measurement volume. Relative to the jet forcing parameters in \citet{Bang_2023}, we decreased both the mean on- and off-times while increasing the mean fraction of pumps that are on at any given time. The chosen mean on- and off-times were 1 s and 4 s, respectively, which corresponded to a dimensionless mean on-time of 1.4 and a mean firing-pump fraction of 20\%. This resulted in a homogeneous turbulence region approximately 20 cm wide and 60 cm tall at the tank centre \citep{bang2026thesis}.

We measured three-dimensional (3D) fluid velocities within a $10 \times 10 \times 10\, \textrm{cm}^3$ cubic volume located in the homogeneous turbulence region (figure \ref{fig:ext_setup}). Three high-speed cameras (Phantom VEO 340L) were focused on the measurement region with a reduced sensor area of $1920 \times1600$ px (relative to the native $2560\times1600$ px) to allow for longer-duration recordings. Two cameras were positioned on one side of the measurement volume, while the third was positioned approximately orthogonal to the other two viewing directions to improve 3D reconstruction of particle positions. LED panels located opposite the cameras resulted in backlit images, so that objects within the measurement volume appeared as shadows. The flow was seeded with neutrally buoyant tracer particles (hereafter referred to as \emph{tracers}), consisting of white polyethylene microspheres with diameters of 212--250 $\mu$m (Cospheric LLC). These tracers were larger than 3 px in the images, allowing the tracer centroids to be determined with subpixel accuracy by fitting a Gaussian to the intensity profile. Although the tracer diameter is comparable to the Kolmogorov length scale, tracers of similar relative size to the Kolmogorov scale have previously been shown to successfully measure fluid velocities in turbulent flows \citep{Masuk_2019,Tan_2020,Wang_2025}. 3D reconstruction of tracer positions was achieved by camera calibration performed by moving a 4 cm sphere mounted on linear stages through predetermined locations. The projection matrix for each camera was then determined using singular value decomposition based on the known sphere positions in laboratory coordinates and their corresponding image centroids. The calibration was subsequently refined using volumetric self-calibration \citep[VSC;][]{Wieneke_2008}. VSC was performed by gently stirring the fluid with a low tracer concentration, and matching and triangulating tracer positions from the three cameras. Residual disparities between the detected and reprojected tracer centroids were used to determine local corrections to the projection matrices, and the procedure repeated with progressively finer spatial resolution and smaller convergence thresholds.

Using the calibrated and self-corrected imaging system, we measured the flow field by acquiring a burst of three images every 2 s where the images within each burst were taken every 1 ms. The short inter-frame interval allows the tracer motion to be approximated as linear because it is much shorter than the Kolmogorov time scale of approximately 40 ms, while the longer interval between bursts reduces the temporal correlation of the velocity field between successive bursts. Such data were acquired over 30 minutes from 900 bursts. Tracer positions were reconstructed independently in each frame. Before triangulation, the local VSC disparity correction was applied to the tracer image centroids in each camera view. The corrected centroids were then matched and triangulated, and the reconstructed positions were optimised by minimising the reprojection error. Reconstructions were only accepted when the average residual disparity across the cameras was less than 0.4 px, a commonly used threshold \citep{Wieneke_2008, Wieneke_2018}. For each burst, we formed three-frame tracer trajectories by linking 3D tracer positions using nearest-neighbour matching. Using a linear fit to each trajectory then gave the tracer position vector $\bm{x}_t$ and tracer velocity vector $\bm{u}_t$ at the middle frame, where the subscript $t$ denotes tracer quantities.

We computed the flow characteristics within the measurement volume using the 900 snapshots of velocity field obtained from the tracer velocities. We found the velocity field had a mean $\bm{U} = [0.897, -0.669, 0.397]$ cm s$^{-1}$ and root-mean-square (RMS) values $\bm{u}_{rms} = [3.94, 3.94, 3.22]$ cm s$^{-1}$ in the [$x$, $y$, $z$] directions, respectively. We note the mean velocity is small relative to the RMS velocity fluctuations, showing that the turbulence dominates the velocity field. We also note that the RMS velocities in both horizontal directions are the same and larger than the vertical RMS velocity, which indicate large-scale horizontal isotropy and weak large-scale anisotropy across the horizontal and vertical directions due to the absence of jet forcing in the vertical direction. 

\begin{figure}
    \centering
    \includegraphics[width=0.7\linewidth]{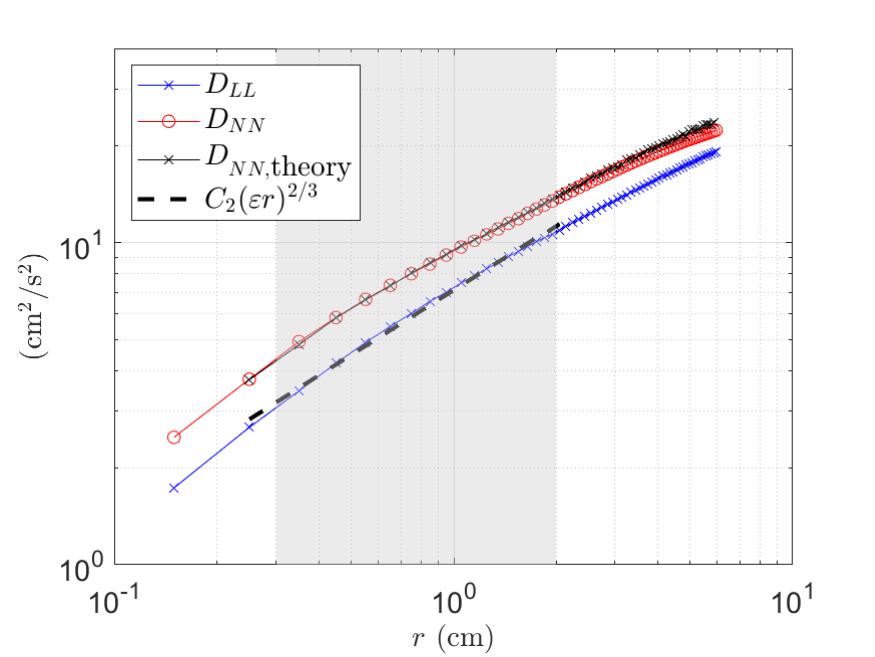}
    \caption{Second-order structure functions, with the inertial subrange indicated by the shaded region.}
    \label{fig:2nd_str_function_Gr2010}
\end{figure}

To examine the turbulence statistics in the inertial subrange, we computed the longitudinal and transverse second-order structure functions, $D_{LL}$ and $D_{NN}$, respectively. Figure \ref{fig:2nd_str_function_Gr2010} shows the data. We estimated the mean turbulent dissipation rate $\varepsilon$ by fitting the longitudinal structure function data to $D_{LL} = C_2 (\varepsilon r)^{2/3}$ over the inertial subrange ($0.3 \le r \le 2$ cm) where $C_2=2.0$ and $r$ is the separation distance (dashed black line in figure \ref{fig:2nd_str_function_Gr2010}). For homogeneous, locally isotropic turbulence, $D_{NN}$ is uniquely determined by $D_{LL}$ \citep{Pope_2000} as 
\begin{equation}
    D_{NN}(r) = D_{LL}(r) + \frac{1}{2} r \frac{\partial}{\partial r} D_{LL}(r).
    \label{eq:D_NN_theory}
\end{equation}
The transverse structure function calculated from this relation ($D_{NN,\text{theory}}$) agrees closely with the measured $D_{NN}$ (solid black line with crosses in figure \ref{fig:2nd_str_function_Gr2010}), indicating that the large-scale anisotropy decays to scale-local isotropic turbulence in the inertial subrange.

The resulting turbulent flow characteristics are summarised in table \ref{table:Turbulent flow characteristics}. The large-scale turbulent velocity scale, $u_L$, is computed as the quadratic mean of the three RMS velocity components. The integral length and time scale are estimated as $L = 0.5 u_L^3/\varepsilon$ and $T = L/u_L$, respectively. Taylor microscale is also computed by $\lambda = u_L({30\nu/\varepsilon})^{1/2}$ using the small-scale isotropic relation \citep{Pope_2000}. These expressions have previously been shown to provide reliable estimates for the scales of turbulent motions in this facility \citep{Bang_2023}. The Reynolds numbers based on the integral length scale and Taylor microscale are also calculated as $Re_L = u_L L/\nu$ and $Re_\lambda = u_L \lambda/\nu$, respectively. The small value of $Fr_\eta$ indicates that the characteristic small-scale turbulent acceleration is much weaker than gravity, though this is not necessarily an instructive measure of the strength of the small-scale turbulence since gravity does not play a role in homogeneous density flows. Note that the measurement and analysis procedures described above yielded turbulence characteristics consistent with those obtained from 2D particle image velocimetry data collected under identical jet forcing conditions \citep{bang2026thesis}.

\begin{table}
    \begin{center}
\def~{\hphantom{0}}
    \begin{tabular}{ccccccccc}
    $u_L$ (cm/s) & $L$ (cm)& $T$ (s)& $\varepsilon$ ($\textrm{cm}^2\, \textrm{s}^{-3}$) & $\eta$ (mm) & $\tau_\eta$ (ms) & $Re_\lambda$ & $Re_L$ & $Fr_\eta$\\
    3.72 & 3.82 & 1.03& 6.72 & 0.203 & 39.5 & 286 & 1357 & 0.013
    \end{tabular}
    \caption{Turbulent flow characteristics}
    \label{table:Turbulent flow characteristics}
    \end{center}
\end{table}

\subsection{Finite-size spherical particles}
\label{sec:finite-size spherical particles}

To span a broad range of the parameter space governing the behaviour of particle settling in turbulence, we use thirteen different finite-size, negatively buoyant spherical particles (hereafter referred to as \emph{particles}). Table \ref{table:Particle properties} summarises their physical properties. The list includes particles of seven different materials that span material densities of $1.02 \le \rho_p \le 2.7$ $\textrm{g}\, \textrm{cm}^{-3}$: acetal resin (AR), nylon (NY), polytetrafluoroethylene (PTFE), polyethylene (PE), polystyrene (PS), soda lime glass (SL), and aluminium (AL). The list also includes particles spanning an order of magnitude in size with diameters of $0.50 \le d_p \le 6.35$ $\textrm{mm}$. All particles were purchased from commercial suppliers (AR1--2, NY1--3, PTFE, and AL from McMaster-Carr; PE1--3, PS, and SL from Cospheric LLC; and AR3 from  Ball and Roller Store) and stored in water to allow water absorption to reach equilibrium and ensure that their densities remained stable during experiments.

\begin{table}
    \begin{center}
\def~{\hphantom{0}}
    \begin{tabular}{cccccccc}
     Particle & $d_p$ (mm) & $\rho_p$ ($\textrm{g}\, \textrm{cm}^{-3}$) & $v_s$ (cm s$^{-1}$) &  $\tau_{p,SN}$ (s) & $d_p/v_s$ (s) & $\gamma$ & $Re_{p,t}$  \\
    AR1 & 2.38 $\pm$ 0.03 & 1.42 $\pm$ 0.02 & 13.6 $\pm$ 0.1 &  0.0645 & 0.0176 & 1.42 & 308 \\
    AR2 & 1.59 $\pm$ 0.02 & 1.42 $\pm$ 0.02 & 8.38 $\pm$ 0.1 &  0.0482 & 0.0190 & 1.42 & 127\\
    AR3 & 1.00 $\pm$ 0.02 $^{\ast}$ & 1.40 $\pm$ 0.02 $^{\ast}$ & 5.81 $\pm$ 0.2 &  0.0292 & 0.0172 & 1.40 & 56\\
    NY1 & 6.35 $\pm$ 0.05 & 1.14 $\pm$ 0.01 & 15.4 $\pm$ 0.2 &  0.195 & 0.0414 & 1.14 & 932 \\
    NY2 & 3.18 $\pm$ 0.03 & 1.14 $\pm$ 0.01 & 8.79 $\pm$ 0.1 &  0.107 & 0.0361 & 1.14 & 267\\
    NY3 & 1.59 $\pm$ 0.02 & 1.14 $\pm$ 0.01 & 5.00 $\pm$ 0.1 &  0.0543 & 0.0318 & 1.14 & 76 \\
    PTFE & 1.59 $\pm$ 0.02 & 2.12 $\pm$ 0.03 & 19.2 $\pm$ 0.1 &  0.0405 & 0.0083 & 2.12 & 291\\
    PE1 & $0.85 - 1.00\, ^{\ast\ast}$ & 1.02 $\pm$ 0.01 $^{\ast}$ & 0.90 $\pm$ 0.1 & 0.0419 & 0.103 & 1.02 & 8\\
    PE2 & $0.85 - 1.00\, ^{\ast\ast}$ & 1.10 $\pm$ 0.03 $^{\ast}$ & 2.42 $\pm$ 0.1 & 0.0319 & 0.0382 & 1.10 & 21\\
    PE3 & $0.85 - 1.00\, ^{\ast\ast}$ & 1.25 $\pm$ 0.05 $^{\ast}$ & 3.80 $\pm$ 0.1 & 0.0290 & 0.0244 & 1.25 & 34\\
    PS & 6.3 $\pm$ 0.05 & 1.05 $\pm$ 0.01 & 10.3 $\pm$ 0.1  & 0.235 & 0.0610 & 1.05 & 622\\
    SL & $0.50 - 0.60\, ^{\ast\ast}$ & 2.5 $\pm$ 0.05 $^{\ast}$ & 8.23 $\pm$ 0.3 & 0.0157 & 0.0067 & 2.5 & 43\\
    AL & 0.79 $\pm$ 0.01 $^{\ast}$ & 2.7 $\pm$ 0.2 $^{\ast}$ & 13.6 $\pm$ 0.1 & 0.0225& 0.0058 & 2.7 & 103\\
    \end{tabular}
    \caption{Particle properties. Uncertainty bounds represent 95 \% confidence intervals obtained using the bootstrap method, while $^{\ast}$ and $^{\ast\ast}$ denote manufacturer-specified values and size ranges, respectively.}
    \label{table:Particle properties}
    \end{center}
\end{table}

The particle diameter and density data were obtained either from direct measurements or from manufacturer specifications, depending on the particle type. For particles larger than 1 mm, the diameter $d_p$ and mass were measured using a calliper and a precision balance, respectively, after surface water was gently removed. For AR3 and AL, however, their small size made complete removal of surface water difficult. We therefore relied on manufacturer specifications for their diameters and densities. We also used manufacturer-specified ranges for diameters and densities for PE1--3 and SL since these particles were supplied with specified ranges in their physical properties. For the calculation of the derived particle properties, the midpoint of each manufacturer-specified diameter range was used as the representative diameter.

The terminal settling velocity $v_s$ in quiescent water was measured within the same measurement volume in the tank as used for the fluid flow measurements. For each particle type, 10--15 individual settling trajectories were recorded. The mean settling velocity was first obtained for each trajectory, and the terminal settling velocity and its 95 \% confidence interval were then estimated using bootstrapping. Other quantities, such as $Re_{p,t}$, $\tau_{p,SN}$, and $d_p/v_s$, were computed based on these $v_s$ values. While we did observe path instabilities in the form of non-vertical settling trajectories for particles where $Re_{p,t} > 210$ \citep{Horowitz2010effect,Raaghav2022path,Cabrerebooman2024path}, their influence on the mean vertical velocity was small and we concluded that the path instabilities had little influence on the terminal settling velocity estimation. We do not further analyse these path instabilities since complete path-instability cycles could not be reliably resolved in the 10 cm vertical extent of the measurement volume.

\begin{figure}
    \centering
    \includegraphics[width=0.7\linewidth]{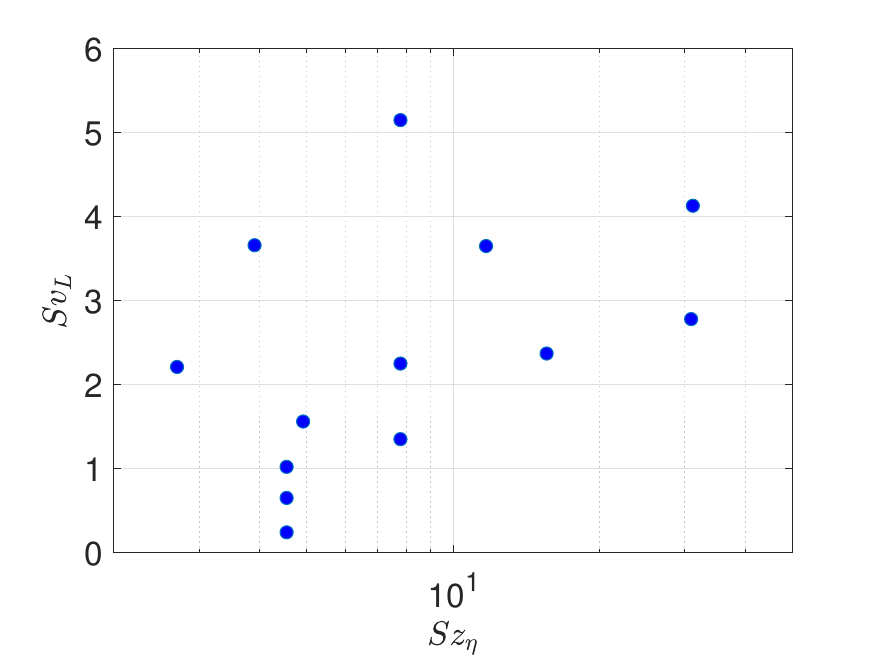}
    \caption{Particles in $Sz_\eta - Sv_L$ space. }
    \label{fig:SZ Sv parameter space}
\end{figure}

The dimensionless parameters characterising the particle-turbulence interactions are summarised in table \ref{table:dimless_parameters}, with figure \ref{fig:SZ Sv parameter space} showing the positions of the particles in $Sz_\eta$--$Sv_L$ space. All particles are finite-size ($Sz_\eta>1$) and span settling numbers on either side of unity ($0.24 < Sv_L < 5.2$). The particle set includes groups with comparable $Sz_\eta$ ($\approx4.5$ and $7.8$) but varying $Sv_L$, and a group with comparable $Sv_L$ ($\approx2.3$) but varying $Sz_\eta$, allowing the effects of settling velocity and particle size to be examined separately. However, the effects of turbulence parameters, such as $Re_\lambda$ and $Fr_\eta$, are not examined because all experiments were conducted under the same statistical turbulent flow conditions.

\begin{table}
    \begin{center}
\def~{\hphantom{0}}
    \begin{tabular}{cccccc}
     Particle & $Sz_\eta$ & $Sv_L$ & $St_{\eta,SN} $ & $St_{\eta,d_p/v_s}$ & $Fr_{\eta,p}$\\
    AR1 & 12 & 3.7 & 1.6 & 0.44 &0.045\\
    AR2 & 7.8 & 2.3 & 1.2 & 0.48 &0.045\\
    AR3 & 4.9 & 1.6 & 0.74 & 0.44 &0.047\\
    NY1 & 31 & 4.1 & 4.9 & 1.0 &0.11\\
    NY2 & 16 & 2.4 & 2.7 & 0.91 &0.11\\
    NY3 & 7.8 & 1.4 & 1.4 & 0.80 &0.11\\
    PTFE & 7.8 & 5.2 & 1.0 & 0.21 &0.025\\
    PE1 & 4.5 & 0.24 & 1.1 & 2.6 &0.68\\
    PE2 & 4.5 & 0.65 & 0.81 & 0.97 &0.15\\
    PE3 & 4.5 & 1.0 & 0.73 & 0.62 &0.067\\
    PS & 31 & 2.8 & 5.9 & 1.5 &0.28\\
    SL & 2.7 & 2.2 & 0.40 & 0.17 &0.022\\
    AL & 3.9 & 3.7 & 0.57 & 0.15 &0.021\\
    \end{tabular}
    \caption{Dimensionless parameters for particle-turbulence interaction.}
    \label{table:dimless_parameters}
    \end{center}
\end{table}

\subsection{Simultaneous measurement of particle motion and surrounding flow}
\label{sec:simultaneousmeasurements}

We now describe the experimental procedures and data analysis methods used to track the motion of settling particles while simultaneously measuring the surrounding fluid flow around them. For each particle type, we collected data of particle settling in the low particle concentration regime where particle-particle interactions were negligible. Particles were manually released just below the water surface to prevent air bubbles from becoming trapped on their surfaces. Rapidly settling particles were released individually, and a new particle was introduced only after the previously released particle had passed below approximately half the tank depth. For more slowly settling particles ($Sv_L \lesssim 2$), multiple particles (5--10) were released simultaneously from slightly separated initial positions. The turbulent flow dispersed the particles during settling, and instances in which two particles remained close to one another within the measurement volume were rare.

We used the same three-frame burst imaging method as used in the flow field measurements (\S\ref{sec:turbulence_tank_and_flow_characterisation}), but with the images within each burst taken every 10 ms. This rate balanced the desire for high temporal resolution (approximately four flow velocity snapshots per Kolmogorov timescale) with the need to have sufficiently long recording durations based on the camera memory (allowing the slowly settling particles to traverse the height of the measurement volume). The recorded images were processed to identify the particle image centroids and reconstruct the corresponding 3D positions. Particle image centroids were first identified in the middle-frame images so that the particle positions were temporally aligned with the fluid velocity measurements made by tracking the tracer particles over the course of each burst. For particles with $d_p>1.5$ mm, centroids were detected using the MATLAB function \textit{imfindcircles}. For particles with $d_p<1.5$ mm, intensity-weighted centroids were determined using \textit{regionprops}, because their sizes were smaller than the minimum circle-search radius required by \textit{imfindcircles}. The centroid locations identified in the three camera views were then matched and triangulated, and the reconstructed particle positions were optimised by minimising the reprojection error. This procedure produced raw 3D particle trajectories at a temporal resolution of 100 Hz. For each particle type, we collected data of 100--118 trajectories.

To reduce measurement noise and improve the estimation of particle velocity and acceleration, we fitted a second-order polynomial over a sliding window of five consecutive raw particle positions. The fitted trajectories yielded the particle position $\bm{x}_p$, velocity $\bm{v}$, and acceleration $\bm{a}$. The five-point window spanned 40 ms, approximately equal to the Kolmogorov timescale $\tau_\eta$. This choice provided the best balance between noise reduction and avoiding excessive attenuation of the real acceleration fluctuations when compared with higher polynomial orders, longer fitting windows, and Gaussian-kernel filtering \citep[see][for further details]{bang2026thesis}. 

From the tracer velocity data, we computed two coarse-grained fluid velocities, $\tilde{\bm{u}}$ and $\overline{\bm{u}}$. From the tracer positions $\bm{x}_t$ and velocities $\bm{u}_t$, $\tilde{\bm{u}}$ was estimated as the coarse-grained fluid velocity found by spatially averaging the tracer velocities within a search radius $D_s/2$ centred on the particle: 
\begin{equation}
    \tilde{\bm{u}}(\bm{x}_p)=\frac{1}{N_s}\sum_{i\in\mathcal{S}}\bm{u}_{t,i},\qquad \mathcal{S}=\left\{i:\frac{d_p}{2}\leq\left|\bm{x}_p-\bm{x}_{t,i}\right|\leq\frac{D_s}{2}\right\}.
\label{eq:estimation of fluid velocity}
\end{equation}
We used $D_s = 30$ mm, which is approximately 80\% of the integral length scale value and yielded enough tracers $N_s$ to calculate a reliable mean (approximately 15--25 per particle observation). To reduce truncation of the search volume near the measurement boundaries, the analysis was restricted to instances when the particle centre was at least 1 cm away from the upper and lower boundaries of the measurement volume. We interpret $\tilde{\bm{u}}$ as the local coarse-grained fluid velocity surrounding the particle that resolves motions at or larger than the integral scale of turbulence but averages out smaller scale turbulent motions.

We constructed $\overline{\boldsymbol{u}}$ to estimate the large-scale background flow. This coarse-grained velocity field is computed from spatially averaging the tracer velocities over the entire measurement volume at each time step during particle settling. We assumed that the background flow had a spatial correlation length that scaled with the tank size and was therefore much larger than the dimensions of the measurement volume. Thus, spatial averaging over the entire measurement volume would smooth out any turbulent velocity fluctuations and provide a representative estimate of the large-scale background flow at any given time.

\subsection{Computing statistics from particle trajectories with unequal lengths}
\label{sec:Estimation of statistics from trajectories with unequal lengths}

Computing statistics from particle trajectories with unequal lengths requires careful treatment. If all data across trajectories are pooled together, longer trajectories would have a greater influence. To avoid this bias, we implemented a method where each trajectory was treated as an independent realisation and assigned equal weight. 

For each quantity of interest, including particle velocity, particle acceleration, and surrounding fluid velocity, an empirical probability density function (PDF) was first computed separately for each trajectory using a common set of bins. We then computed an overall PDF by averaging across trajectories. The mean, standard deviation, skewness, and kurtosis were computed from this trajectory-averaged PDF. The uncertainty in the statistical moments arising from the finite number of trajectories was quantified using a bootstrap procedure, where the PDFs from different trajectories were sampled with replacement and averaged to obtain an ensemble of trajectory-averaged PDFs. Computing the statistical moments across the ensemble of PDFs allowed 95 \% confidence intervals to be obtained. The common bins were selected at a sufficiently fine resolution such that the resulting statistics and their 95 \% confidence intervals were insensitive to further refinement of the bin resolution.

An underlying assumption in this approach is that the data from even the shortest trajectories adequately sample the underlying particle velocity distribution. Because the settling velocity varied across different particles, we used the trajectory arc length estimated by summing the lengths of segments for each time step to assess whether a trajectory was long enough to be included in the analysis. The median trajectory length exceeded 8 cm for all particles, which is comparable to or greater than the vertical extent of the measurement volume. A small number of trajectories, primarily for slowly settling particles, were shorter than 5 cm, but these cases were rare and generally remained within the measurement volume for more than half the integral time scale, $0.5T$. Their influence on the reported statistics is therefore expected to be limited.

\section{Results}
\label{sec:Results}

\begin{figure}
    \centering
    \includegraphics[width=\linewidth]{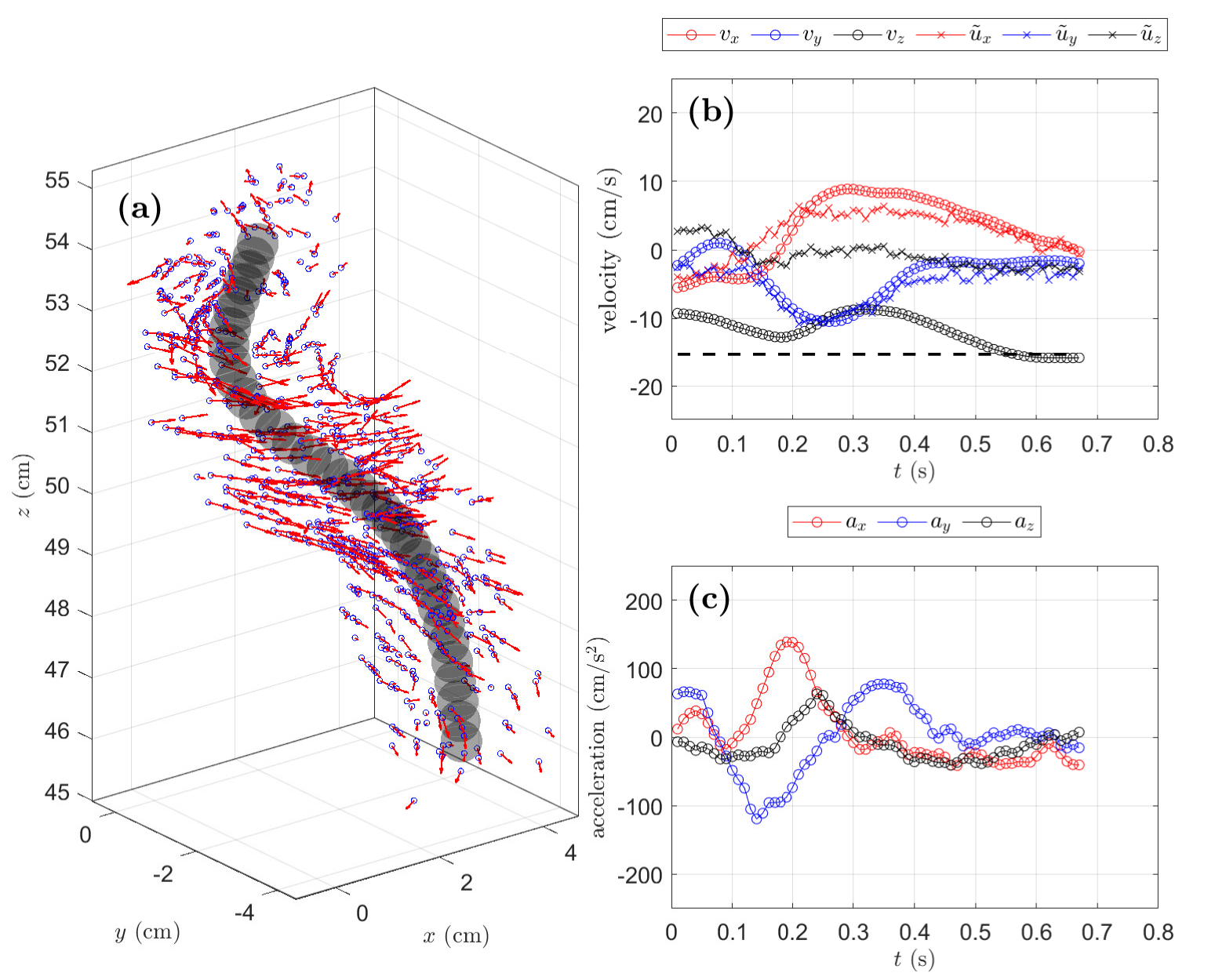}
    \caption{Representative trajectory and corresponding data for the NY1 particle. (a) Particle trajectory is represented by transparent black spheres, which are drawn to scale ($d_p = 6.35$ mm) for every other sample for clarity. Blue circles and red arrows indicate the tracer positions and velocities, respectively. (b) Time series of particle and surrounding fluid velocities, where the black dashed line indicates the terminal settling velocity in quiescent water. (c) Particle acceleration time series.}
    \label{fig:representative_NY1_high_acc}
\end{figure}

Figure \ref{fig:representative_NY1_high_acc} shows data from an example trajectory of the NY1 particle, which has the largest $Sz_\eta$ and $Re_{p,t}$ values of all particles. In panel (a), we observe that during the part of the trajectory where there is a large horizontal fluid flow in the $+x$ and $-y$ directions, the particle trajectory responds via a pronounced horizontal displacement in the same direction. And when the horizontal flow subsequently weakens, the particle trajectory becomes nearly vertical. This behaviour is reflected in the time series of velocities and accelerations shown in panels (b) and (c). The particle accelerates horizontally in response to the strong horizontal flow it encounters during $0.05 \le t \le 0.2$ s. There are also vertical variations in particle velocity despite relatively little change in the vertical fluid velocity in its surroundings during $0.25 \le t \le 0.5$ s. These observations demonstrate the type of insights that can be gained from time-resolved simultaneous measurements of particle and surrounding fluid motion. Below, we gather such data from many particle trajectories for all particles to examine particle acceleration statistics (\S\ref{sec:Particle acceleration}) and particle velocity statistics (\S\ref{sec:Particle_velocity}).

\subsection{Particle acceleration}
\label{sec:Particle acceleration}

\begin{figure}
    \centering
    \includegraphics[width=1\linewidth]{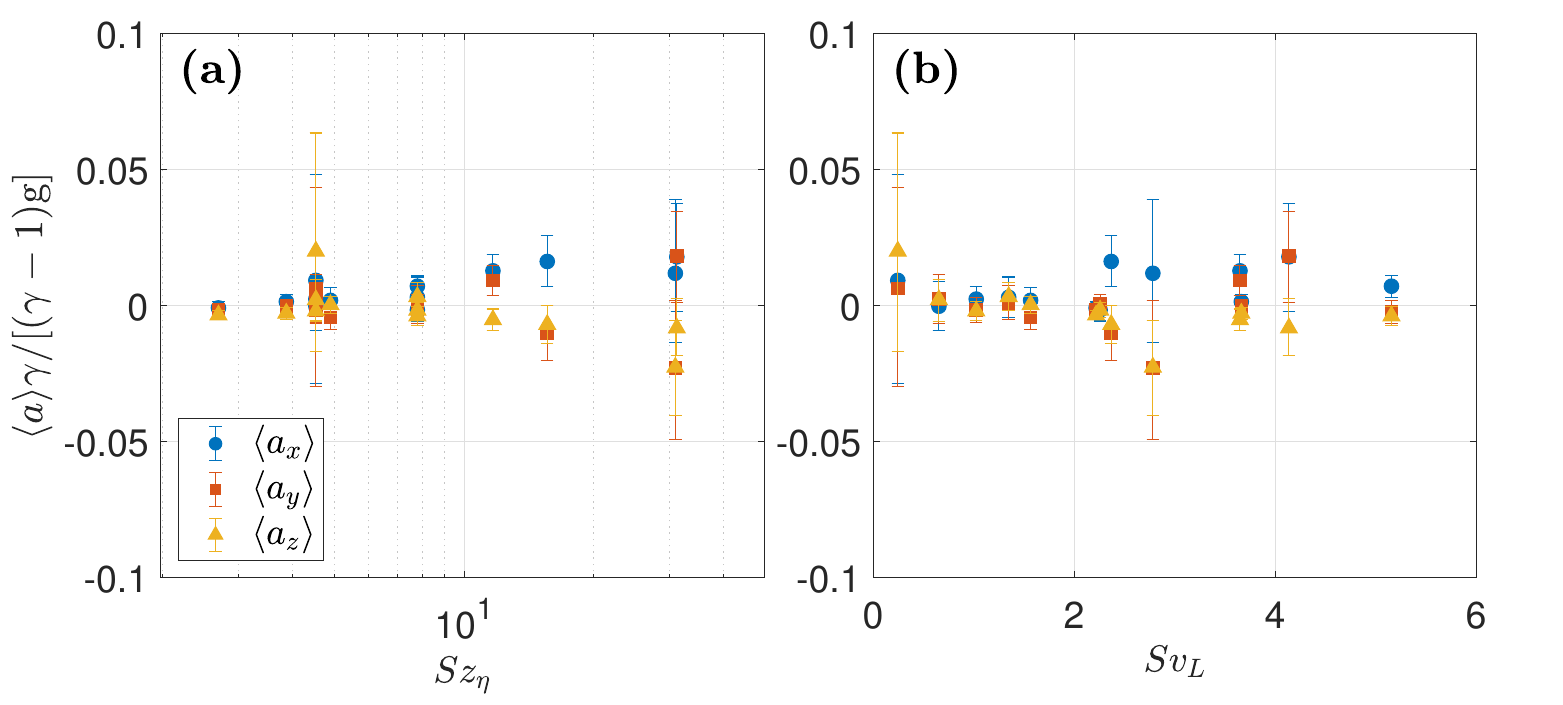}
    \caption{Mean particle acceleration normalised by the submerged-weight acceleration as functions of (a) $Sz_\eta$ and (b) $Sv_L$. Error bars indicate 95 $\%$ confidence intervals obtained by bootstrapping.}
    \label{fig:mean_acc_Sz_Sv}
\end{figure}

Introducing the decomposition $\bm{a} = \langle \bm{a} \rangle + \bm{a}^\prime$ into Eq.~\eqref{eq:particle_acceleration}, we can write the mean and fluctuating components of the particle acceleration as
\begin{subequations}
    \begin{align}
        \langle\bm{a}\rangle&= -\left(\frac{\gamma - 1}{\gamma}\right)\textrm{g} \bm{e}_z + \langle\bm{a}_{\textrm{hd}}\rangle\label{eq:mean_acc}\\
        \bm{a}^\prime &= \bm{a}_{\textrm{hd}}^\prime\label{eq:fluct_acc}.
    \end{align}
\end{subequations}

\begin{figure}
    \centering
    \includegraphics[width=0.85\linewidth]{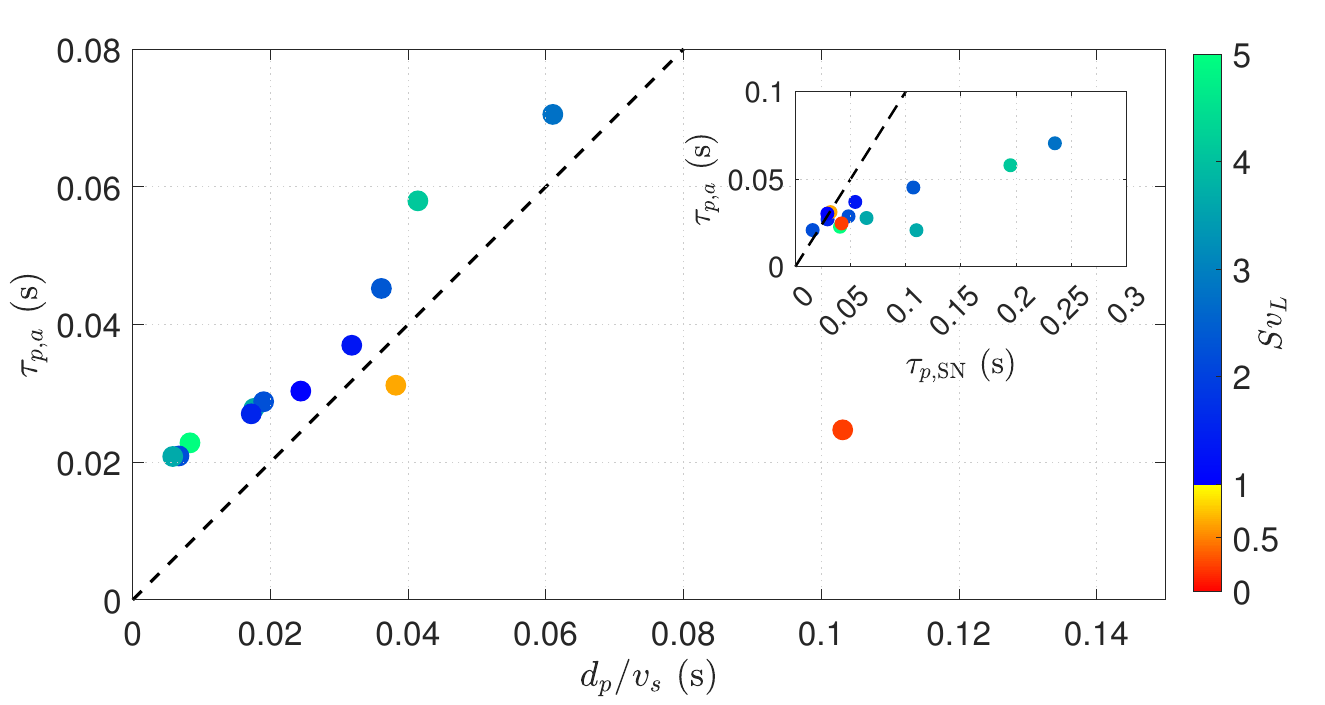}
    \caption{Particle acceleration timescale as a function of the predicted particle timescales $d_p/v_s$ (main panel) and $\tau_{p,\textrm{SN}}$ (inset). For visual clarity, separate colour maps are used for particles with $Sv_L<1$ and $Sv_L>1$. The black dashed lines denote one-to-one agreement between the measured and predicted timescales.}
    \label{fig:tau_p_a_vs_tau_p_est}
\end{figure}

Figure \ref{fig:mean_acc_Sz_Sv} shows the dimensionless mean particle acceleration, which is found to be close to zero for all particles in all three directions. From Eq.~\eqref{eq:mean_acc}, this indicates that the particle weight, buoyancy, and hydrodynamic force balance on average, and that $\langle\bm{a}_{\textrm{hd}}\rangle \approx \left[({\gamma - 1})\textrm{g}/{\gamma}\right] \bm{e}_z$.

We now turn to examine the particle acceleration timescale, which characterises the persistence of the force experienced by finite-size particles as they settle in turbulence. It is quantified from the particle acceleration autocorrelation, computed using the procedure described in Appendix \ref{app:autocorrelation}. Figure \ref{fig:tau_p_a_vs_tau_p_est} shows this particle acceleration timescale, $\tau_{p,a}$, against two candidate timescales, $\tau_{p,SN}$ and $d_p/v_s$. Because the particle acceleration timescales are similar in all three directions, $\tau_{p,a}$ is defined as their RMS value. Separate colour maps distinguish particles with $Sv_L<1$ and $Sv_L>1$. For weak settling ($Sv_L<1$), we see that $\tau_{p,SN}$ provides a closer estimate of $\tau_{p,a}$ than $d_p/v_s$. In contrast, $d_p/v_s$ provides a closer estimate for a wide range of values for $Sv_L>1$. In this regime, $\tau_{p,a}$ is only slightly greater than $d_p/v_s$, whereas $\tau_{p,SN}$ can over-predict the measured timescale by up to a factor of three with no trend with respect to $Sv_L$. Thus, the particle acceleration timescale seems to shift from $\tau_{p,SN}$ under weak settling to $d_p/v_s$ when settling is dominant.

\begin{figure}
    \centering
    \includegraphics[width=\linewidth]{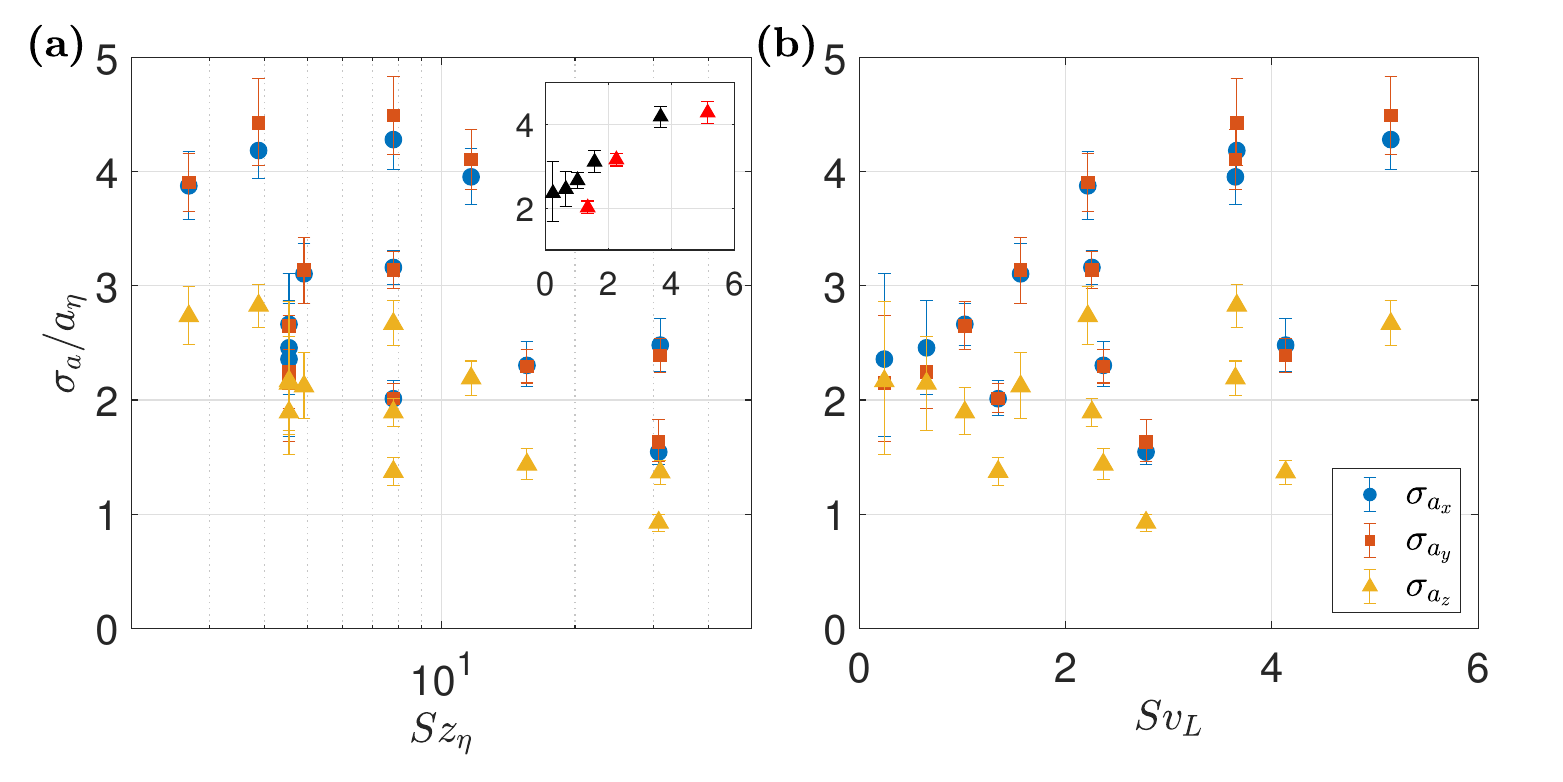}
    \caption{Standard deviations of particle acceleration normalised by Kolmogorov acceleration scale as functions of (a) $Sz_\eta$ and (b) $Sv_L$. Error bars indicate 95 $\%$ confidence intervals obtained by bootstrapping. The inset in panel (a) shows $\sigma_{a_x}/a_\eta$ for particles with $Sz_\eta \approx 4.5$ (black markers, AR3, PE1-3, and AL) and for $Sz_\eta \approx 7.8$ (red markers, AR2, NY3, and PTFE) as a function of $Sv_L$.}
    \label{fig:rms_prtc_acc_Sz_Sv}
\end{figure}

For $Sv_L>1$ particles, the observation $\tau_{p,a} \approx d_p/v_s$ can be interpreted by considering their interaction with turbulent eddies along settling-dominated trajectories. For an eddy of size $\ell$, the settling-based crossing time is $\ell/v_s$ whereas the eddy turnover time is $\ell/u_\ell$ with $u_\ell$ being the characteristic velocity at scale $\ell$. The shorter of these two timescales provides the characteristic interaction time between the particle and the eddy. Since $u_\ell \leq u_L$ for $\ell \leq L$, the condition $Sv_L>1$ ensures that $\ell/v_s$ is the shorter timescale. That means that the particle always crosses an eddy before that eddy evolves to a new state, for eddies across the full spectrum of turbulence ($\eta \leq \ell \leq L$). Now, if the interaction timescale is small compared to the particle timescale, \textit{i.e.} $\ell/v_s \ll \tau_{p,a}$, the particle may not fully adjust its motion to the eddy-scale fluid velocity before leaving the eddy, while the opposite is true for $\ell/v_s \gg \tau_{p,a}$. With $\tau_{p,a} \approx d_p/v_s$, these two regimes correspond to $\ell \ll d_p$ and $\ell \gg d_p$, respectively. Thus, the data suggest that the particle diameter defines a characteristic spatial scale separating turbulent motions to which a settling particle does and does not respond. It does not respond to eddies $\ell < d_p$, while it does respond to eddies $\ell > d_p$. This applies for scale $\ell$ across all turbulent length scales when $Sv_L>1$. 

For neutrally buoyant finite-size particles, the acceleration response timescale is known to scale as the turnover time for an eddy of the same size as the particle \citep{voth2002measurement}, $\tau_{p,a}\sim(d_p^2/\varepsilon)^{1/3}$ \citep{Qureshi_2007, VOLK_2011}. Thus, the particle only responds to eddies of size $\ell > d_p $. Despite the different interaction timescales, both settling and neutrally buoyant finite-size particles exhibit similar spatial filtering, responding to turbulent eddies larger than their size while filtering out eddies smaller than their size.

This understanding can be extended to predict an inertial range scaling for the particle acceleration variance for finite-sized settling-dominant particles, where the particle size falls in the inertial range $\eta \ll d_p \ll L$ and the settling velocity is larger than the turbulent fluctuations $Sv_L > 1$. In this regime, the particle acceleration variance is expected to depend on the turbulent energy dissipation rate $\varepsilon$ based on Kolmogorov\textquotesingle s second similarity hypothesis \citep{Pope_2000} and on the particle size and settling velocity via the timescale $d_p/v_s$ to account for spatial filtering and settling. From dimensional arguments, this yields that the particle acceleration variance scales as $\sigma_a^2 \sim \varepsilon v_s / d_p$. Making this scaling relation dimensionless with the Kolmogorov acceleration scale and taking the square root yields 
\begin{equation}
\frac{\sigma_a}{a_\eta} \sim St_{\eta,d_p/v_s}^{-1/2}
\label{eq:particleaccstd_scaling}
\end{equation}
for the particle acceleration standard deviation. This acceleration standard deviation reflects the magnitude of the hydrodynamic force fluctuations per unit particle mass (Eq.~\eqref{eq:fluct_acc}), underscoring its importance.

\begin{figure}
    \centering
    \includegraphics[width=0.8\linewidth]{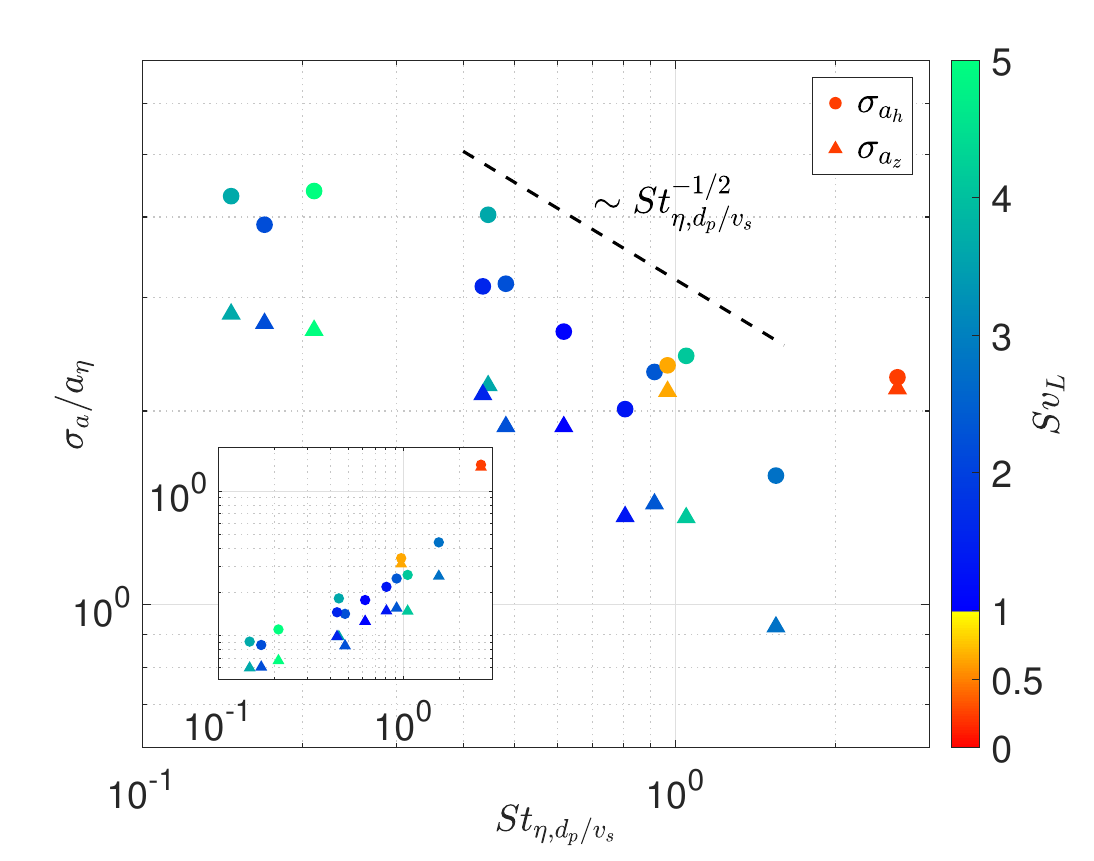}
    \caption{Standard deviations of particle acceleration normalised by the Kolmogorov acceleration scale as a function of $St_{\eta,d_p/v_s}$. The inset shows the same acceleration statistics normalised by the submerged-weight acceleration, $[(\gamma-1)g]/\gamma$. Circles and triangles represent the horizontal and vertical components, respectively. Marker colour indicates $Sv_L$, with separate colour maps used for particles with $Sv_L<1$ and $Sv_L>1$.}
    \label{fig:rms_prtc_acc_St_dpvs}
\end{figure}

Figure \ref{fig:rms_prtc_acc_Sz_Sv} shows the standard deviations of particle acceleration as functions of $Sz_\eta$ and $Sv_L$. Unlike the size dependence observed for neutrally buoyant particles \citep{voth2002measurement, Qureshi_2007, VOLK_2011}, we see no clear trend in the acceleration standard deviations with $Sz_\eta$. Examining particles within two groups of similar size ($Sz_\eta \approx 4.5$ and $Sz_\eta \approx 7.8$) in the inset, particles with greater $Sv_L$ generally exhibit greater acceleration fluctuations. Also, among particles with similar settling velocities ($Sv_L \approx 2.3$), smaller particles exhibit larger acceleration standard deviations (not shown). However, there is no clear dependence with respect to $Sv_L$ for all particles. 

Figure \ref{fig:rms_prtc_acc_St_dpvs} tests the scaling in Eq.~\eqref{eq:particleaccstd_scaling}, showing the standard deviation of the particle accelerations normalised by the Kolmogorov-scale acceleration as a function of the Stokes number based on the settling timescale. Separate colour maps are used for particles with  $Sv_L<1$ and $Sv_L>1$ and the two horizontal components being very similar in magnitude (figure \ref{fig:rms_prtc_acc_Sz_Sv}) are combined into an overall horizontal component as $\sigma_{a_h}=[(\sigma_{a_x}^2+\sigma_{a_y}^2)/2]^{1/2}$. For $Sv_L>1$ particles, both acceleration components appear to be consistent with the scaling in Eq.~\eqref{eq:particleaccstd_scaling} for $St_{\eta,d_p/v_s} > 0.3$. The saturation observed for small $St_{\eta,d_p/v_s}$ may be related to the fact that for small enough $d_p/v_s$, the turbulent eddies are effectively frozen from the particle\textquotesingle s perspective and/or the particle is sufficiently small that there is effectively no spatial filtering. In both scenarios, no further gains in particle acceleration are expected to occur, leading to saturation of its standard deviation. A similar saturation of the particle acceleration variance below a size threshold was previously observed for neutrally buoyant particles \citep{voth2002measurement,Brown_2009,Bourgoin_2011}. However, unlike previous particle acceleration variance data for sub-Kolmogorov particles, no data collapse is observed when the acceleration standard deviations are plotted against $St_{\eta,SN}$ (not shown), further confirming the scaling in Eq.~\eqref{eq:particleaccstd_scaling} and the accompanying physical picture that particle size acts as a spatial filter and particle settling velocity sets the encounter rate of new turbulent eddies via the crossing-trajectories mechanism \citep{Yudine59}. Appendix \ref{app:Influence of settling on particle acceleration} analyses particle trajectories directly to provide further support for this crossing-trajectories effect for $Sv_L>1$ particles.

For $Sv_L<1$ particles (PE1 and PE2), figure \ref{fig:rms_prtc_acc_St_dpvs} shows the acceleration fluctuation magnitudes are similar across particles and directions. From the fact that these particles are also the same size, these results suggest that $Sv_L < 1$ particles can be approximated as neutrally buoyant for predicting the particle acceleration variance. 

The inset in figure \ref{fig:rms_prtc_acc_St_dpvs} shows the same acceleration standard deviations, but normalised by the submerged-weight acceleration. From a dynamical perspective, the acceleration fluctuations normalised in this way can be interpreted as the standard deviation of the hydrodynamic force as a fraction of the mean vertical hydrodynamic force, $\langle {\bm{a}_{\textrm{hd}}^\prime}^2 \rangle ^{1/2} /  \langle \bm{a}_{\textrm{hd},z} \rangle$. The increase with $St_{\eta,d_p/v_s}$ is because the submerged-weight acceleration decreases faster than the particle acceleration fluctuations, showing that the standard deviation of the hydrodynamic force increases as a fraction of the mean (vertical) hydrodynamic force with increasing $d_p/v_s$. While the data collapse to some degree on these axes, we do not believe there is a meaningful scaling since it implies a scaling between $Fr_{\eta,p}$ and $St_{\eta,d_p/v_s}$ that is not fully tested since we do not vary the turbulent dissipation rate.

Lastly in figure \ref{fig:rms_prtc_acc_St_dpvs}, the horizontal component can be observed to be greater than the vertical component, despite the small-scale isotropy of the turbulence. This acceleration anisotropy may arise in part from the continuity effect, in which rapidly settling particles experience shorter correlation times for the horizontal fluid velocity than for the vertical fluid velocity, resulting in faster variations in the sampled horizontal fluid velocity and hence larger horizontal acceleration fluctuations \citep{Csanady63}. But, the stronger horizontal velocity fluctuations at large scales related to the jet forcing could also be enhancing this effect. 

\begin{figure}
    \centering
    \includegraphics[width=\linewidth]{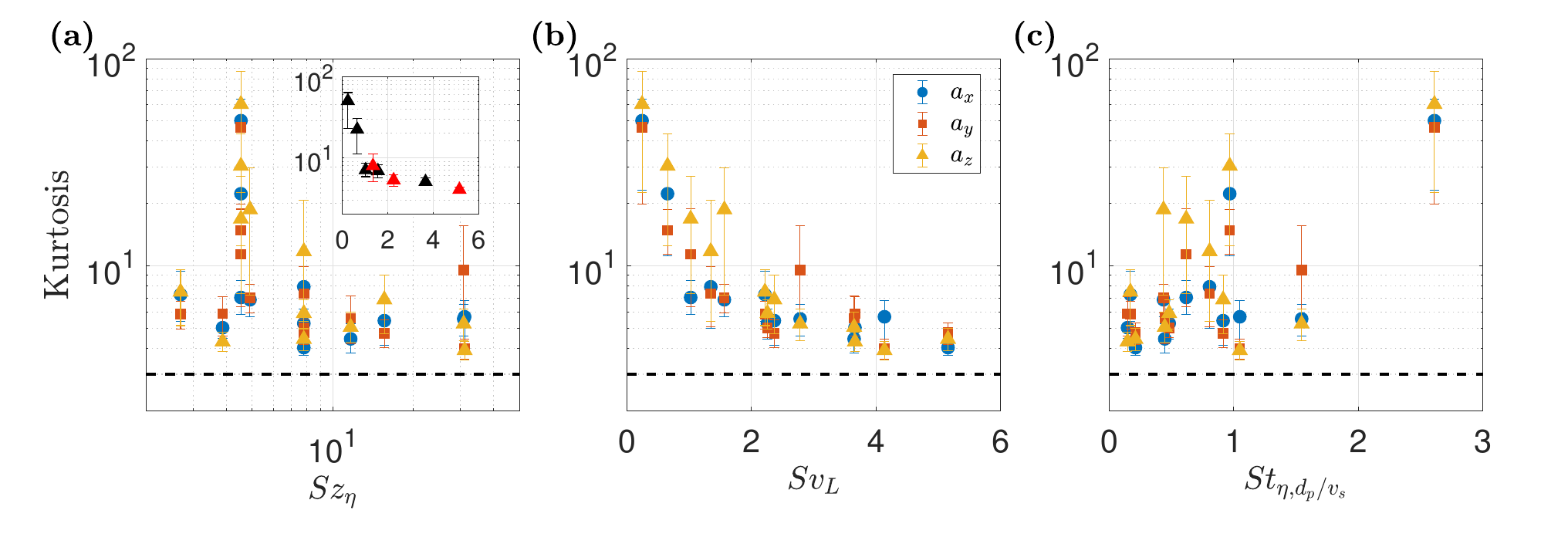}
    \caption{Kurtosis of particle acceleration as a function of (a) $Sz_\eta$, (b) $Sv_L$, and (c) $
    St_{\eta,d_p/v_s}$. Error bars indicate 95 $\%$ confidence intervals obtained by bootstrapping. The horizontal black dashed line indicates the Gaussian value of kurtosis, 3. The inset in panel (a) shows kurtosis of $a_x$ for particles with $Sz_\eta \approx 4.5$ (black markers, AR3, PE1-3, and AL) and for $Sz_\eta \approx 7.8$ (red markers, AR2, NY3, and PTFE) as a function of $Sv_L$.}
    \label{fig:kurt_acc_Sz_Sv_St_dpvs}
\end{figure}

Figure \ref{fig:kurt_acc_Sz_Sv_St_dpvs} shows the particle acceleration kurtosis as a function of $Sz_\eta$, $Sv_L$, and $St_{\eta,d_p/v_s}$ in order to examine the intermittency of particle accelerations. All acceleration kurtosis values exceed 3, indicating heavier-than-Gaussian tails. In contrast to the size dependence reported for neutrally buoyant particles \citep{VOLK_2011}, no clear trend is observed with $Sz_\eta$ (panel (a)). However, within the two similar-size groups ($Sz_\eta \approx 4.5$ and $ Sz_\eta \approx 7.8$), faster-falling particles tend to have smaller kurtosis values (panel (a) inset). The same tendency is evident in panel (b), where slower-falling particles exhibit larger kurtosis values and, therefore, more intermittent acceleration fluctuations. In particular, the three particles with similar settling velocities ($Sv_L \approx 2.3$; AR2, NY2, and SL) exhibit comparable kurtosis values despite their different sizes. And unlike the acceleration standard deviation, the kurtosis shows no clear dependence on $St_{\eta,d_p/v_s}$ (panel (c)). To better understand the dependency of particle acceleration kurtosis on its settling velocity, examples of particle trajectories during extreme acceleration events are given in Appendix \ref{app: extreme_acc_trajectories}. They show that the similarly sized PE1 ($Sv_L<1$) and AL ($Sv_L>1$) particles that have different settling velocities exhibit very distinct trajectories during extreme acceleration events. While the PE1 particle trajectory during an extreme acceleration event suggests interactions with a particle-scale turbulent vortex, the trajectory of the AL particle during an extreme acceleration event is more associated with adjustment to new turbulent eddies encountered during settling. In other words, extreme accelerations for particles with $Sv_L>1$ are more closely associated with the crossing-trajectories effect (Appendix \ref{app:Influence of settling on particle acceleration}) and less associated with spatial filtering as is the case for neutrally buoyant particles \citep{VOLK_2011, Lalescu18}. 

\subsection{Particle velocity and settling modulation}
\label{sec:Particle_velocity}

\begin{figure}
    \centering
    \includegraphics[width=0.8\linewidth]{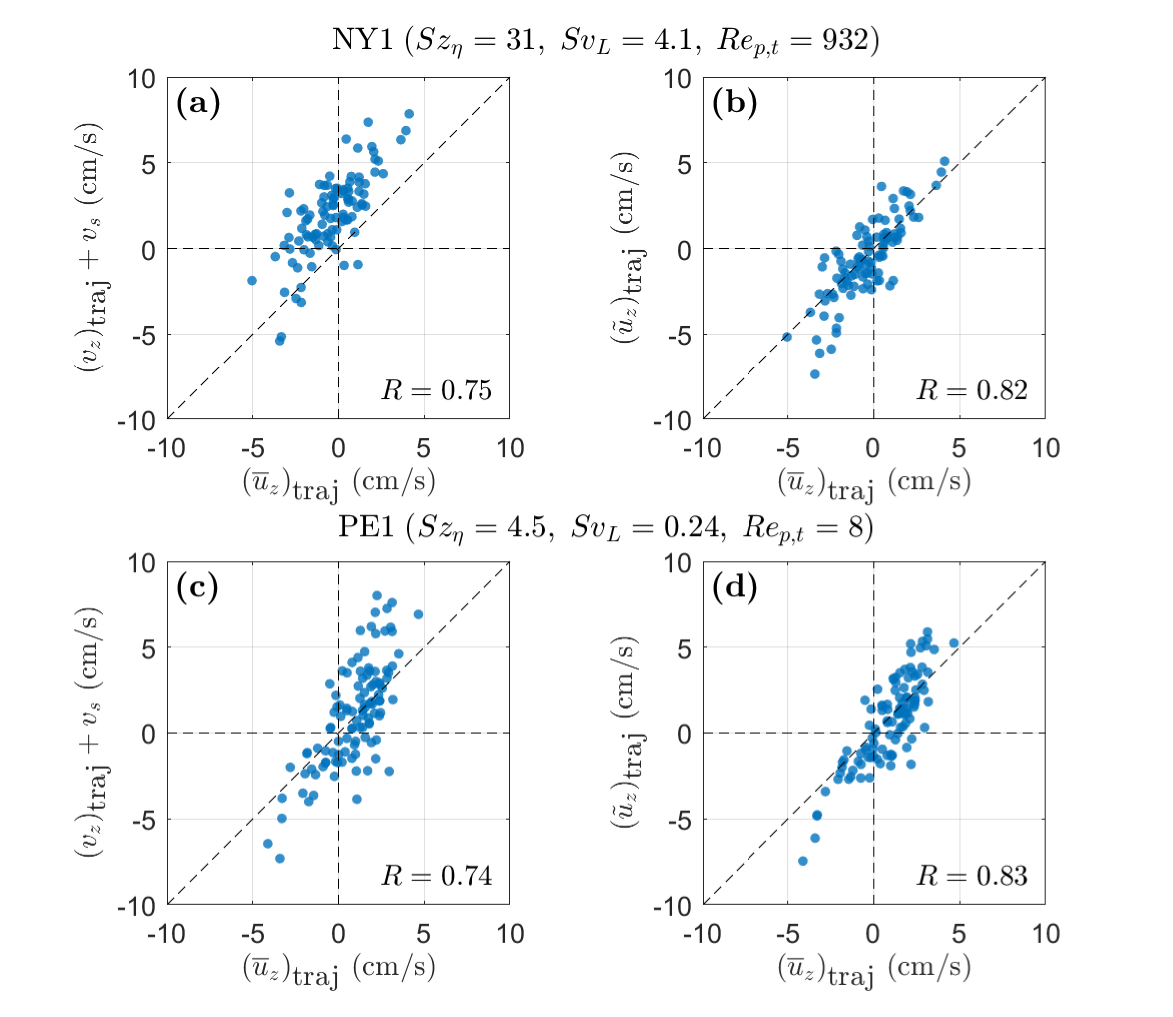}
    \caption{Trajectory-averaged vertical velocities for NY1 (a,b) and PE1 (c,d). Comparisons between (a,c) settling-corrected particle and background flow velocities and (b,d) surrounding integral-scale fluid and background flow velocities. Each marker represents an individual trajectory. Dashed diagonal lines indicate the one-to-one relationship, and $R$ denotes the Pearson correlation coefficient. }
    \label{fig:net_settling_modulation_NY1_PE1}
\end{figure}

\begin{figure}
    \centering
    \includegraphics[width=\linewidth]{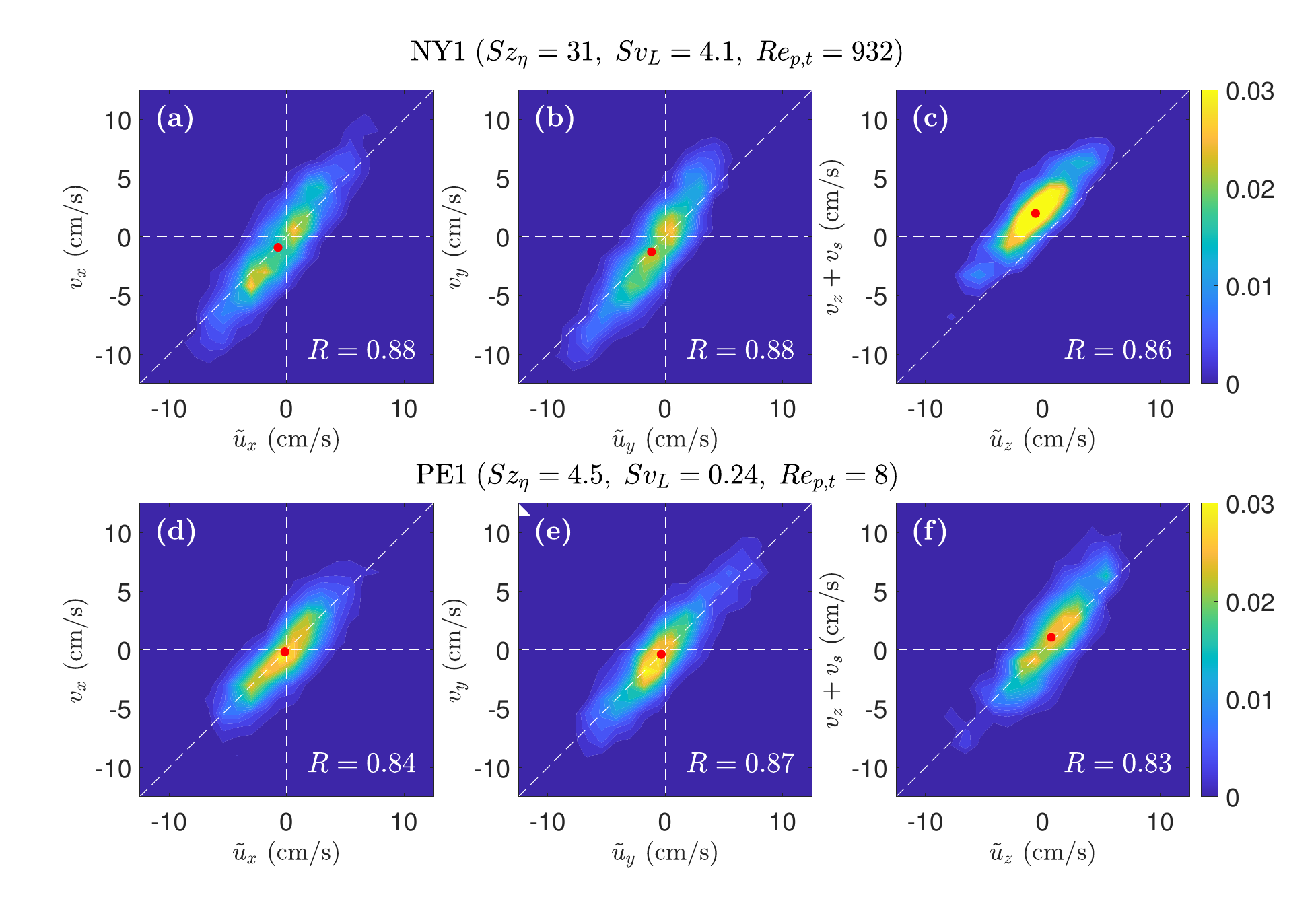}
    \caption{Joint probability density functions of particle velocity and surrounding fluid velocity for NY1 and PE1 particles in the directions $x$ (a,d), $y$ (b,e), and $z$ (c,f). In panels (c,f), the particle vertical velocity is corrected by the terminal settling velocity. The white diagonal dashed line indicates the one-to-one relation between the two variables and the red dot indicates their joint mean. The Pearson correlation coefficient $R$ is shown in the lower right corner of each panel.}
    \label{fig:joint_pdf_vu_NY1_PE1}
\end{figure}

We start by demonstrating the importance of the background flow in the definition of the net settling modulation $\Delta_{NS}$ in Eq.~\eqref{eq:net_settling_quantification}. Figure \ref{fig:net_settling_modulation_NY1_PE1} shows the trajectory-averaged values of the settling-corrected particle vertical velocity $\left(v_z + v_s\right)_{\text{traj}}$, the average fluid velocity in the measurement volume $\left(\overline{u_z}\right)_{\text{traj}}$, and the local coarse-grained fluid velocity around the particle $\left(\tilde{u}_z\right)_{\text{traj}}$, plotted against each other for the NY1 and PE1 particles as representative examples of the full dataset. We see that the measurement-volume-averaged fluid velocity has a clear influence on both the particle velocity and the local fluid velocity, even when averaged over trajectories. Separate analysis (not shown) also shows that $\overline{u_z}$ is also reasonably steady over the duration of particle trajectories for all but the slowest settling particles. Thus, the data show that what may appear as variability in the particle settling rate in $(v_z + v_s)$ or variability in the local velocity sampled by the particle in $\tilde{u}_z$ are actually caused by background flows at scales larger than the turbulent integral scale that persist for longer than the integral timescale. This highlights the importance of using $\langle v_z - \overline{u}_z + v_s \rangle$ and $\langle \tilde{u}_z - \overline{u}_z \rangle$, rather than $\langle v_z - \tilde{u}_z + v_s \rangle$ and $\langle \tilde{u}_z \rangle$, to quantify net settling modulation and preferential sampling, respectively, in Eq.~\eqref{eq:net_settling_quantification}. 

Next, we examine how the particle velocity relates to the local flow coarse-grained at the integral scale. Figure \ref{fig:joint_pdf_vu_NY1_PE1} shows the joint PDFs of the settling-corrected particle velocity, $\bm{v}+v_s\bm{e}_z$, where $\bm{e}_z$ denotes the unit vector in the upward vertical direction, and the local fluid velocity coarse-grained at the integral scale, $\tilde{\bm{u}}$. These joint PDFs were obtained by averaging the joint PDFs calculated from individual trajectories with equal weighting. The data are for the NY1 and PE1 particles, which are again chosen as representative examples. We observe that the joint PDFs are peaked close to the one-to-one line for both particles and in all three directions, with the notable exception of the vertical component of the NY1 particle where the joint PDF is displaced vertically upwards from the 1:1 line indicating a reduced settling relative to the local flow. These observations hold across all particles despite variations in particle size and settling velocity. Across the full dataset of all particles and for all components, the correlation coefficients between $\bm{v}+v_s\bm{e}_z$ and $\tilde{\bm{u}}$ range from $0.8$ to $0.9$ while the slopes from linear regressions range from $0.95$ to $1.15$ with a median value of $1.05$. Thus, at the simplest approximation, particles appear to follow the local coarse-grained fluid velocity while settling at a constant rate that may or may not be altered from its terminal settling velocity in quiescent fluid.

This behaviour can be understood from the particle acceleration correlation timescale analysis in \S\ref{sec:Particle acceleration}. To deduce how finite-sized settling particles will respond to the integral-scale turbulent motions, we can compare the particle timescale $\tau_{p,a}$ with the turbulence integral timescale $T = L / u_L$. For strongly settling particles ($Sv_L>1$), $\tau_{p,a} \approx d_p/v_s$ and its ratio with the integral timescale gives $(d_p/v_s)/(L / u_L) = Sz_L / Sv_L$, where $Sz_L = d_p/L$ is the size number defined with respect to the integral scale. The ratio $Sz_L / Sv_L$ is guaranteed to be below unity for settling-dominated ($Sv_L > 1$) particles that are sub-integral scale in size ($Sz_L < 1$). In our dataset, $(Sz_L / Sv_L) < 1$ even for weakly settling particles ($Sv_L<1$) since only very small particles ($Sz_L \ll 1$) tend to settle slower than the turbulent velocity fluctuations. The condition $(Sz_L / Sv_L) < 1$ suggests particles have sufficient time to respond and come to equilibrium with the integral-scale turbulent eddies, which then explains why particles follow the local coarse-grained fluid motion while settling.

The deviation of the particle velocity from $\bm{v} = \tilde{\bm{u}} - v_s\bm{e}_z$, where $v_s$ is the terminal settling velocity in quiescent fluid, captures how turbulence alters settling relative to the local flow and how particle and fluid inertia may cause relative motion with respect to the local flow. To facilitate a quantitative analysis, we define the settling-corrected relative velocity
\begin{equation}
    \bm{v}_r \equiv \bm{v} - \tilde{\bm{u}}+ v_s \bm{e}_z,
    \label{eq:kinematic_model_particle_velo}
\end{equation}
and note that it can be interpreted as a `coarse-grained slip velocity' in the horizontal directions and as a `settling-corrected coarse-grained slip velocity' in the vertical direction. We then decompose it into a mean and fluctuations about the mean  $\bm{v}_r = \langle \bm{v}_r \rangle + \bm{v}_r'$ where $\langle \cdot \rangle$ denotes the ensemble mean obtained using the procedure described in \S\ref{sec:Estimation of statistics from trajectories with unequal lengths}. The mean and fluctuating components are given by
\begin{subequations}
    \begin{align}
        \langle \bm{v}_r \rangle &= \langle \bm{v} \rangle - \langle \tilde{\bm{u}} \rangle +v_s \bm{e}_z,
        \label{eq:rel_velo_mean} \\
        \bm{v}_r'  &=  \bm{v}' -  \tilde{\bm{u}}'.
        \label{eq:rel_velo_fluct}
    \end{align}
\end{subequations}
The spectra of the relative velocity fluctuation do not show any settling-related peaks, implying that $\bm{v}_r'$ only captures the settling-corrected particle-turbulence relative motion as intended. In Appendix \ref{app:Particle_induced_flow_modulation}, we assess how the particle-induced flow modulation affects $\tilde{\bm{u}}$, finding that it is expected to have a negligible influence for most particles. In Appendix \ref{app:v_u_mean_std}, we assess the particle\textquotesingle s inertial response to the local flow, focusing on the aspects not related to settling modulation. We find that, apart from the settling behaviour, the particles are effectively inertialess with respect to the local flow coarse-grained at the integral scale. 

\begin{figure}
    \centering
    \includegraphics[width=1\linewidth]{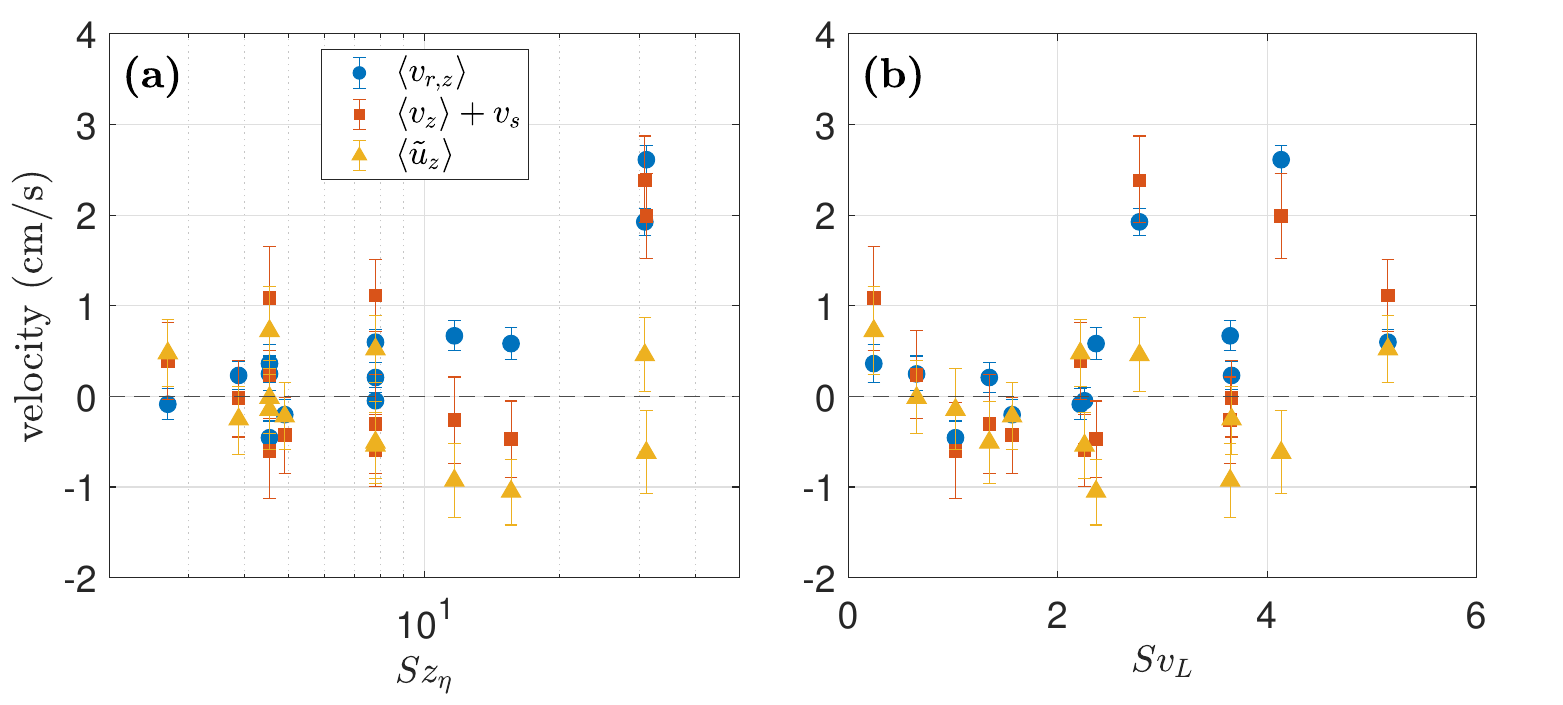}
    \caption{Mean vertical velocities, $\langle v_{r,z}\rangle$, $\langle v_z\rangle+v_s$, and $\langle \tilde{u}_z\rangle$, as functions of (a) $Sz_\eta$ and (b) $Sv_L$. Error bars indicate 95 \% confidence intervals obtained by bootstrapping.}
    \label{fig:vertical_velocities}
\end{figure}

Focusing on the settling modulation, we note that the vertical component of Eq.~\eqref{eq:rel_velo_mean} is exactly the first term on the right-hand side of Eq.~\eqref{eq:net_settling_quantification}, the part of the net settling modulation $\Delta_{NS}$ that captures how turbulence might affect the mean hydrodynamic force on the particle and alter the settling velocity relative to the local flow. Figure \ref{fig:vertical_velocities} shows this term as well as the other terms from the vertical component of Eq.~\eqref{eq:rel_velo_mean} as functions of $Sz_\eta$ and $Sv_L$: the mean settling-corrected relative vertical velocity $\langle v_{r,z} \rangle$, the mean settling-corrected particle vertical velocity $(\langle v_{z} \rangle + v_s)$, and the mean local fluid vertical velocity sampled by particles $\langle \tilde{u}_{z} \rangle$. While $\langle v_{r,z} \rangle$ deviates from zero for several particles indicating altered settling, this tendency does not show a clear dependence on either $Sz_\eta$ or $Sv_L$, showing that neither of these two parameters accurately predicts the turbulence-induced settling modulation relative to the local flow. Additionally, we do not see any clear trends in how $(\langle v_{z} \rangle + v_s)$ and $\langle \tilde{u}_{z} \rangle$ vary with $Sz_\eta$ and $Sv_L$. 

\begin{figure}
    \centering
    \includegraphics[width=0.75\linewidth]{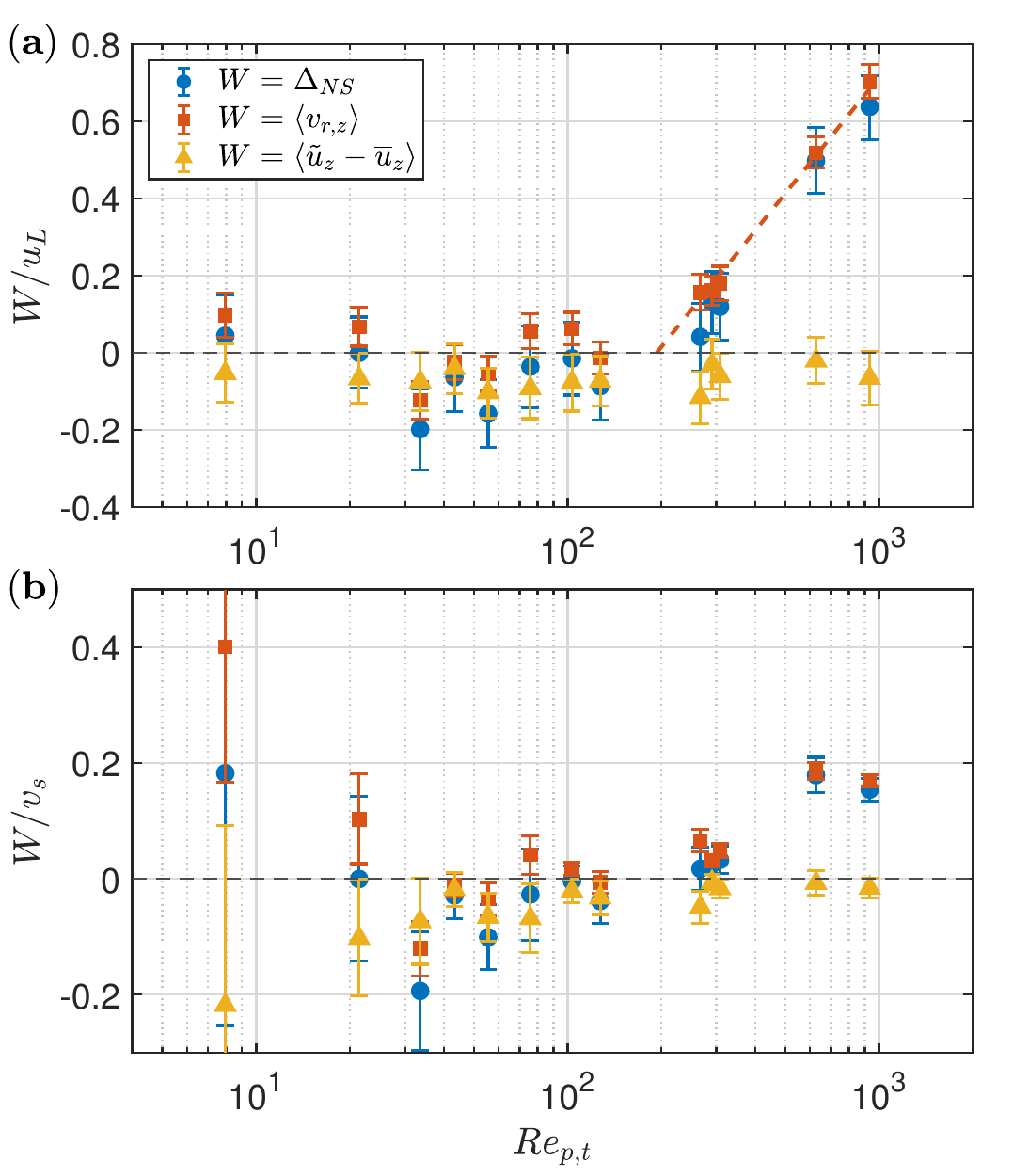}
    \caption{Net settling modulation and its two contributions defined in Eq. \eqref{eq:net_settling_quantification} as functions of $Re_{p,t}$, normalised by (a) $u_L$ and (b) $v_s$. Error bars indicate 95 \% confidence intervals obtained by bootstrapping. The red dashed line in panel (a) indicates a logarithmic fit to the data for $Re_{p,t} \gtrsim 200$: $\langle v_{r,z}\rangle/u_L = \log_{10}(Re_{p,t}/193)$.} 
    \label{fig:net_settling_modulation_Rept}
\end{figure}

However, we see a clearer trend when $\langle v_{r,z} \rangle$ is plotted as a function of the terminal velocity particle Reynolds number $Re_{p,t}$ in figure \ref{fig:net_settling_modulation_Rept}. Panel (a) shows the data made dimensionless by the turbulence velocity scale $u_L$ whereas panel (b) shows the data made dimensionless by the particle terminal settling velocity in quiescent water $v_s$. Positive values signify reduced settling and negative values enhanced settling. We observe that the relative vertical velocity $\langle v_{r,z} \rangle$ shows reduced settling for $Re_{p,t}\lesssim 20$ and $Re_{p,t}\gtrsim 200$, with that settling reduction increasing logarithmically with $Re_{p,t}$ as $\langle v_{r,z}\rangle/u_L = \log_{10}(Re_{p,t}/193)$ for $Re_{p,t}\gtrsim 200$. In between, for $20 \lesssim Re_{p,t} \lesssim 200$, there is a mixture of enhanced and reduced settling with perhaps a weak trend suggesting enhancement to reduction with increasing $Re_{p,t}$. It is notable that the $Re_{p,t}$ regimes that correspond to reduced settling relative to the local flow correspond approximately to two distinct qualitative regimes of the particle\textquotesingle s wake when settling in quiescent fluid. $Re_{p,t}\lesssim 20$ corresponds to attached flow with no particle wake in quiescent fluid whereas $Re_{p,t}\gtrsim 200$ corresponds to particles with non-axisymmetric and unsteady wakes, which are also associated with path instabilities in quiescent fluid. 

Figure \ref{fig:net_settling_modulation_Rept} also shows the preferential sampling $\langle \tilde{u}_z - \overline{u}_z \rangle$ of the net settling modulation. The data suggest that this is near-zero for all particles, with perhaps a slight bias towards preferential sampling of downward flows at the lower end of the range $20 \lesssim Re_{p,t} \lesssim 200$. The near-zero preferential sampling is consistent with the picture that all particles in our dataset have a timescale that is small enough for them to respond to and reach equilibrium with the local flow coarse-grained at the integral scale. 

We conclude that settling modulation in turbulence for our dataset is dominated by dynamic effects that alter the mean hydrodynamic force on the particle (first term on the right-hand side of Eq.~\eqref{eq:net_settling_quantification}) rather than kinematic effects that bias the particle to upward or downward flow (second term on the right side of Eq.~\eqref{eq:net_settling_quantification}). In Appendix \ref{app:Fornari_drag_analysis}, we analyse the changes in the mean hydrodynamic force following \citet{Fornari_2016_reduced} who decompose the mean vertical hydrodynamic force into `steady' components of mean settling, cross-flow-induced drag and nonlinear drag, and a `unsteady' component that absorbs the difference between the known mean vertical force and the modelled `steady' components. However, we find that the data imply negative values of the `unsteady' component for $Re_{p,t} \lesssim 100$ particles, suggesting that the quasi-steady assumption for modelling the `steady' components may not be valid. For $Re_{p,t} \gtrsim 100$ particles, the `unsteady term' is near-zero or positive, but the data imply that the cross-flow and nonlinear terms are negligibly small, and therefore the analysis also does not yield any insights for this set of particles.

To explain the reduced relative settling for $Re_{p,t}\gtrsim 200$ particles, we instead turn to how turbulence may induce changes to the particle wake, and consequently, to the mean vertical hydrodynamic force. This interpretation is supported by the data showing the mean particle acceleration remains close to zero (\S\ref{sec:Particle acceleration}) and that the drag coefficient implied by the instantaneous vertical force and particle Reynolds number lies above the Schiller--Naumann drag correlation (Appendix~\ref{app:Fornari_drag_analysis}). Further, the observed reduction occurs above a particle Reynolds number threshold that is similar to that for path instabilities in quiescent fluid which are driven by unsteady and non-axisymmetric wakes ($Re_{p,t} > 210$). Previous studies of the flow around finite-size spheres held fixed relative to an inflow provide insight into how wake modifications due to turbulence could produce the reduced mean particle-fluid relative settling velocity we observe. For particles whose wakes are intrinsically unsteady and non-axisymmetric under laminar inflow, \citet{Homann2013JFM} showed that upstream turbulence modifies vortex shedding and shortens the mean wake. These changes reduce the downstream surface pressure while leaving the upstream pressure nearly unchanged, thereby increasing the pressure contribution to the mean drag. They also confirmed that the mean drag was greater for turbulent inflow compared to laminar inflow at the same mean flow velocity. A similar turbulence-induced amplification of the pressure drag was also identified by \citet{Peng_2023}. Although these studies considered particles held fixed relative to a turbulent inflow rather than freely settling particles in near-zero-mean-flow turbulence, their findings suggest that turbulence-induced wake modifications could increase the mean upward hydrodynamic force for particles above the $Re_{p,t}$ threshold for unsteady and non-axisymmetric wakes. Consistent with this picture, we observe non-axisymmetric wake structures for the $Re_{p,t} \gtrsim 200$ particles settling in turbulence (Appendix \ref{app:Particle_induced_flow_modulation}).

We also highlight that the wake modification and settling reduction are likely due to turbulent eddies at scales comparable to or slightly greater than the particle size. For inertial spheres settling in surface gravity waves where the flow is characterised by a single length scale much larger than the particle diameter \citep{Bang_2025,Baek_2026_wave}, it was found that the wave-induced flow had little-to-no effect on the spectral characteristics of the path instabilities of settling particles. This clearly suggested that the wake is not affected by the presence of an external flow if there is a clear scale separation between the particle and the flow. 

Finally, we note the importance of properly accounting for unequal trajectories for the net settling modulation (\S\ref{sec:Background_Net_settling_modulation}). If all samples are pooled together instead of following the method in \S\ref{sec:Estimation of statistics from trajectories with unequal lengths} to ensure all trajectories are weighted equally, the mean particle ($\langle v_z\rangle$) and sampled fluid ($\langle \tilde{u}_z\rangle$) velocities would be biased in the positive direction by approximately 3--18\% of $v_s$. These biases would then be incorrectly inferred as a settling modulation towards reduced settling and an apparently stronger loitering effect.

\section{Conclusions}
\label{sec:Conclusions}

We have investigated the dynamics of finite-size spherical particles settling in homogeneous turbulence in the regime applicable to natural aquatic environments using simultaneous measurements of particle motion and the surrounding fluid flow. We considered thirteen different spherical particles that are millimetre-scale ($0.5 \leq d_p \leq 6.35$ mm) and marginally denser than the fluid ($1.02 \leq \gamma \leq 2.7$) settling in a homogeneous turbulent water column ($u_L = 3.7$ cm/s, $\varepsilon = 6.7$ cm$^2$s$^{-3}$). The particle sizes are larger than the Kolmogorov scale ($2.7 \leq Sz_\eta \leq 31$) and have terminal settling velocities in quiescent water that are both smaller and larger than the turbulence velocity fluctuations ($0.24 \leq Sv_L \leq 5.2$). The gravitational accelerations are more important than turbulent accelerations for all particles ($Fr_{\eta,p} < 1$).

We find that particle acceleration statistics differ between weakly and strongly settling particles. For particles with $Sv_L>1$, the particle acceleration timescale in turbulence is in agreement with the time the particle takes to settle a distance equal to its own diameter at its terminal settling velocity in quiescent fluid, $d_p/v_s$. This suggests that the response of finite-size settling particles to turbulent eddies is given by a combination of spatial filtering of eddies smaller than their own size and the crossing-trajectories effect where the settling velocity sets the rate at which new eddies are sampled. When this is combined with Kolmogorov\textquotesingle s second similarity hypothesis, the particle acceleration variance is predicted to scale as $\sigma_a^2 \sim \varepsilon v_s/d_p$. The data are consistent with this scaling, except for a saturation at small $d_p/v_s$ values. For particles with $Sv_L < 1$, the particle timescale was better predicted by the Schiller--Naumann drag correlation timescale and the particle acceleration variance was found to be only a function of size and not settling velocity. Both of these observations suggest that $Sv_L < 1$ particles respond to turbulence in a manner similar to finite-sized neutrally buoyant particles. 

We also find that the settling-corrected particle velocity is strongly correlated to the local fluid velocity coarse-grained at the integral scale. This can be understood from scaling arguments showing that for particles smaller than the integral scale ($Sz_L<1$) that settle faster than the turbulent velocity fluctuations ($Sv_L>1$), the particle timescale ($d_p/v_s$) will always be smaller than the integral timescale, giving particles sufficient time to reach equilibrium with the surrounding flow. However, we also find that the mean settling velocity relative to the local flow can be modified by turbulence. This reduced settling was observed for two different regimes of the particle Reynolds number based on particle size and terminal settling velocity in quiescent fluid. For both $Re_{p,t}\lesssim 20$ and $Re_{p,t}\gtrsim 200$, we measure a settling reduction up to 20\% of the terminal settling velocity in quiescent fluid. These Reynolds number thresholds approximately correspond to regimes where the particle has no wake and where the particle develops an unsteady and non-axisymmetric wake, respectively. While the reduced relative settling for $Re_{p,t}\lesssim 20$ was difficult to explain, we find that the reduced relative settling for $Re_{p,t}\gtrsim 200$ was likely due to turbulence induced disturbance of the particle wake that increases the mean drag for the same relative motion. In the $Re_{p,t}\gtrsim 200$ regime, we find the settling reduction to scale with the turbulence velocity fluctuations and show a logarithmic increase with $Re_{p,t}$.

\appendix

\section{Methods for computing autocorrelations and timescales}
\label{app:autocorrelation}

To compute the autocorrelations of relative velocity and particle acceleration, we subtract the mean of each trajectory for the relative velocity, but the global mean across trajectories for the acceleration. The former isolates the temporal fluctuations within individual trajectories by excluding trajectory-to-trajectory variations in the mean relative velocity, whereas the latter retains such variations in the mean acceleration as part of the variability in the net forcing experienced by the particles. For both quantities, the unbiased autocovariance at each lag is computed from all valid pairs within each trajectory and then averaged across trajectories. A common lag range is used, determined by the shortest trajectory in the dataset. The resulting autocovariance is normalised by its zero-lag value, and the corresponding timescales, $\tau_{p,r}$ and $\tau_{p,a}$, are obtained by integrating the autocorrelations from zero lag to their first zero-crossings, determined by linear interpolation between the last positive and first negative values. Note that even for the fastest-decaying acceleration autocorrelations, approximately four to five sampling points are available before the first zero-crossing, suggesting that the estimated timescales are not strongly affected by the finite temporal resolution. 

\begin{figure}
    \centering
    \includegraphics[width=\linewidth]{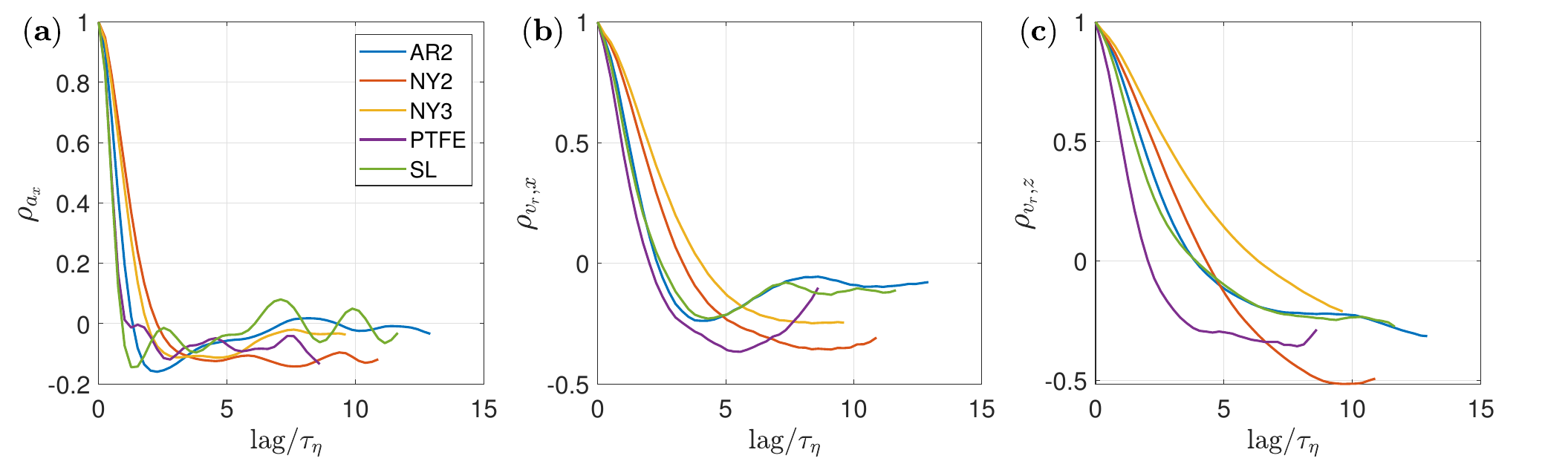}
    \caption{Autocorrelation functions of (a) particle acceleration in the $x$ direction and (b,c) relative velocity in the $x$ and $z$ directions, respectively, for different particles.}
    \label{fig:autocorr_rel_acc}
\end{figure}

Figure \ref{fig:autocorr_rel_acc} shows representative autocorrelation coefficients (autocorrelation functions made dimensionless by the variance) of relative velocity and particle acceleration for five particles. For the relative velocity, the $x$ and $z$ directions are presented, while only the $x$ direction is shown for particle acceleration because of the similarity among the three directions. Both autocorrelations decay from unity and cross zero, followed by weak negative correlations and small secondary oscillations, although particle acceleration decorrelates much more rapidly than relative velocity. For the relative velocity, the initial decay is generally slower in the vertical direction than in the horizontal direction. The autocorrelations decay more rapidly with increasing $Sv_L$ among particles with similar $Sz_\eta \approx 7.8$ (NY3, AR2, and PTFE), whereas the decay becomes slower with increasing $Sz_\eta$ among particles with similar $Sv_L \approx 2.3$ (SL, AR2, and NY2). Similar trends are observed for particle acceleration, although the differences are less pronounced.

\section{Influence of settling on particle acceleration}
\label{app:Influence of settling on particle acceleration}

To demonstrate the crossing-trajectories effect by which settling influences particle-turbulence interactions, we revisit the representative trajectory of the NY1 particle shown in figure \ref{fig:representative_NY1_high_acc}. The magnitude of the particle acceleration peaks as the particle enters and leaves regions of predominantly horizontal flows, suggesting that particle acceleration may depend primarily on the changes in fluid velocity encountered along the settling trajectory.

To characterise the fluid motion that the particle may have recently encountered and is likely to encounter as it settles, we estimate the fluid velocities above and below the particle, $\tilde{\bm{u}}_A$ and $\tilde{\bm{u}}_B$, respectively, using the velocity of tracers within the search volume around the particle. (As described in \S\ref{sec:simultaneousmeasurements}, the search volume is defined by a sphere of diameter $D_s = 30$ mm centred on the particle, which is approximately 80\% of the integral length scale and contains approximately 15-25 tracers.) The tracers are divided into two groups, above and below, based on their positions relative to the particle centre, and the mean tracer velocity in each group is taken to obtain $\tilde{\bm{u}}_A$ and $\tilde{\bm{u}}_B$, respectively.

\begin{figure}
    \centering
    \includegraphics[width=\linewidth]{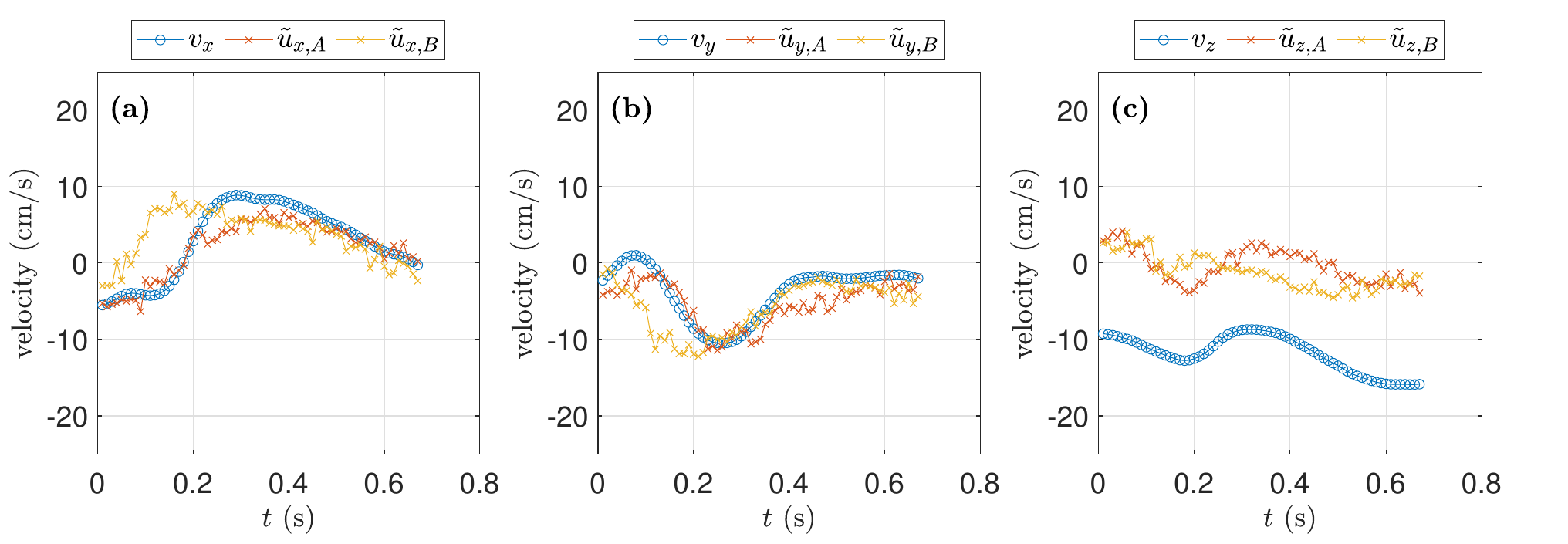}
    \caption{Particle velocity and fluid velocities $\tilde{u}_A$ and $\tilde{u}_B$, computed from tracers located above and below the particle, respectively, in the (a) $x$, (b) $y$, and (c) $z$ directions. Data are taken from figure \ref{fig:representative_NY1_high_acc}.}
    \label{fig:uba_vs_v}
\end{figure}

Figure \ref{fig:uba_vs_v} shows the particle velocity and the fluid velocities above and below the particle along the representative trajectory. In the horizontal directions, the below-particle fluid velocities exhibit substantial changes during $0 \leq t \leq 0.2$ s, followed by similar changes in the corresponding above-particle velocities. Thus, a flow region that is initially located below the particle is sampled by the particle, and then later lies above it. This sequence is consistent with the particle settling through a spatially varying flow that evolves slowly relative to the particle travel time through the flow. In our data, the horizontal components of $\tilde{\bm{u}}_A$ and $\tilde{\bm{u}}_B$ reach magnitudes of approximately $3u_L$, which is in contrast to the vertical components that remain comparatively weak. The largest changes in the horizontal particle velocity, associated with large acceleration events, appear to occur after a clear difference develops between the below- and above-particle fluid velocities.

To quantify this difference and link it to the particle acceleration, we introduce the below-to-above fluid velocity difference, $\Delta \tilde{\bm{u}}_{BA} \equiv (\tilde{\bm{u}}_B-\tilde{\bm{u}}_A)$. For a settling particle, this quantity represents the difference between the fluid motion likely to be encountered and that recently encountered along its trajectory. It may therefore indicate the direction of the particle acceleration after a finite time lag. For example, when a particle approaches a region of flow in the $+x$ direction while settling, $\tilde{u}_{B,x}>0$ and $\tilde{u}_{A,x}\approx 0$, such that $\Delta \tilde{u}_{BA,x}>0$. As the particle enters this flow, it is expected to accelerate in the $+x$ direction, with $a_x>0$. And, as the particle approaches the lower boundary of this flow, $\tilde{u}_{B,x}\approx 0$ and $\tilde{u}_{A,x}>0$, such that $\Delta \tilde{u}_{BA,x}<0$. As the particle leaves the flow, it is expected to accelerate in the $-x$ direction, with $a_x<0$.

\begin{figure}
    \centering
    \includegraphics[width=1\linewidth]{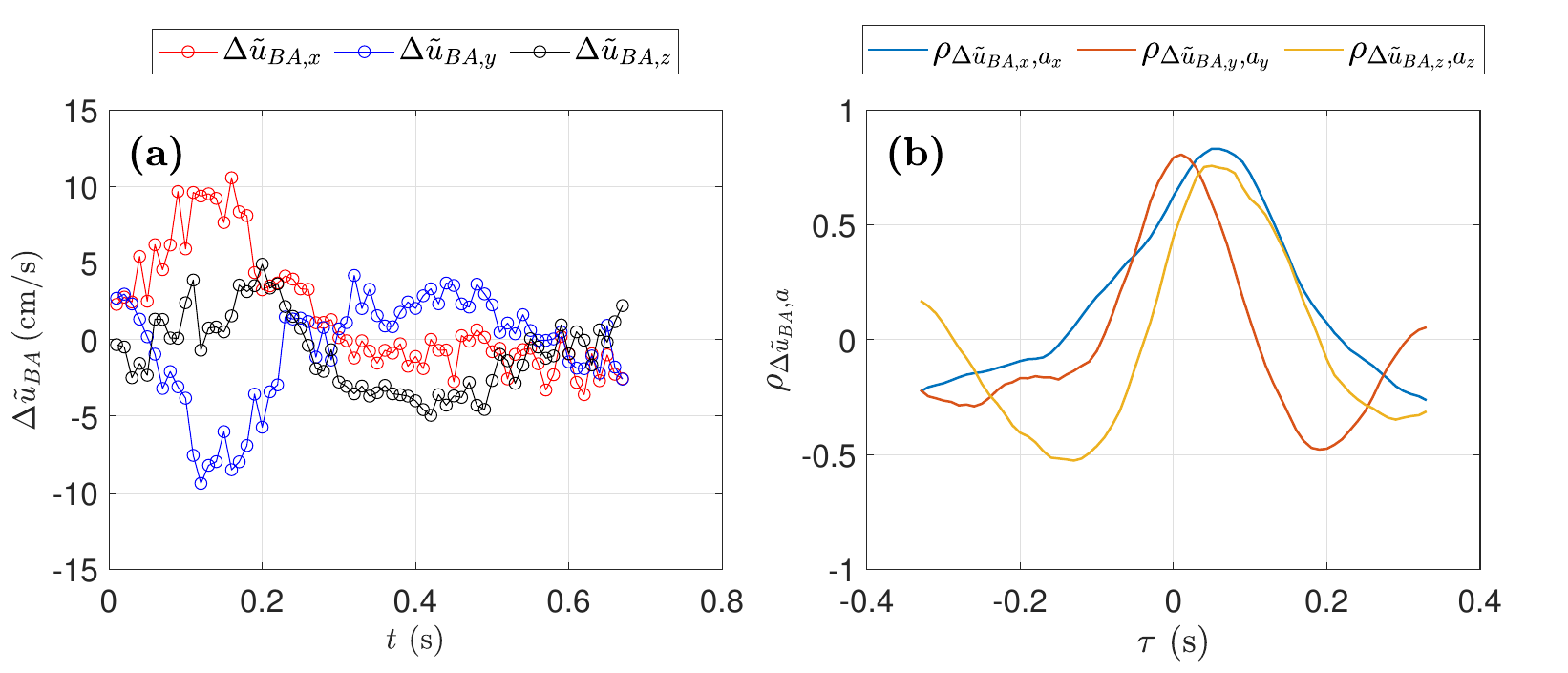}
    \caption{Representative relationship between the below-to-above fluid velocity difference and particle acceleration: (a) below-to-above fluid velocity difference $\Delta  \tilde{\bm{u}}_{BA}$ and (b) normalised cross-correlation between $\Delta  \tilde{\bm{u}}_{BA}$ and particle acceleration $\bm{a}$, computed using the unbiased estimator. Data are taken from figure \ref{fig:representative_NY1_high_acc}.}
    \label{fig:uba_vs_acc}
\end{figure}

Figure \ref{fig:uba_vs_acc} shows this analysis applied to the representative trajectory in Figure \ref{fig:representative_NY1_high_acc}. After a finite time lag, each component of the particle acceleration, displayed in panel (c) in Figure \ref{fig:representative_NY1_high_acc}, appears to vary in the same direction as the corresponding component of $\Delta \tilde{\bm{u}}_{BA}$. This relationship is quantified via the cross-correlation between the two quantities in each direction in panel (b). The cross-correlation is evaluated using an unbiased estimator that accounts for the number of available lagged pairs at each lag and is normalised by the standard deviations of the two signals. A positive lag indicates that the particle acceleration follows the below-to-above fluid velocity difference. The cross-correlation peaks at lags of approximately $0.05$ s in the $x$ and $z$ directions and $0.01$ s in the $y$ direction, with peak values of approximately $0.8$. These results indicate a strong positive correlation between the particle acceleration and the preceding below-to-above fluid velocity difference along this representative trajectory. The finite lag can be interpreted as the time required for the particle to reach the flow region initially located below it or to leave the flow region initially located above it, after which its acceleration begins to change in response to the newly encountered flow.

\begin{figure}
    \centering
    \includegraphics[width=0.9\linewidth]{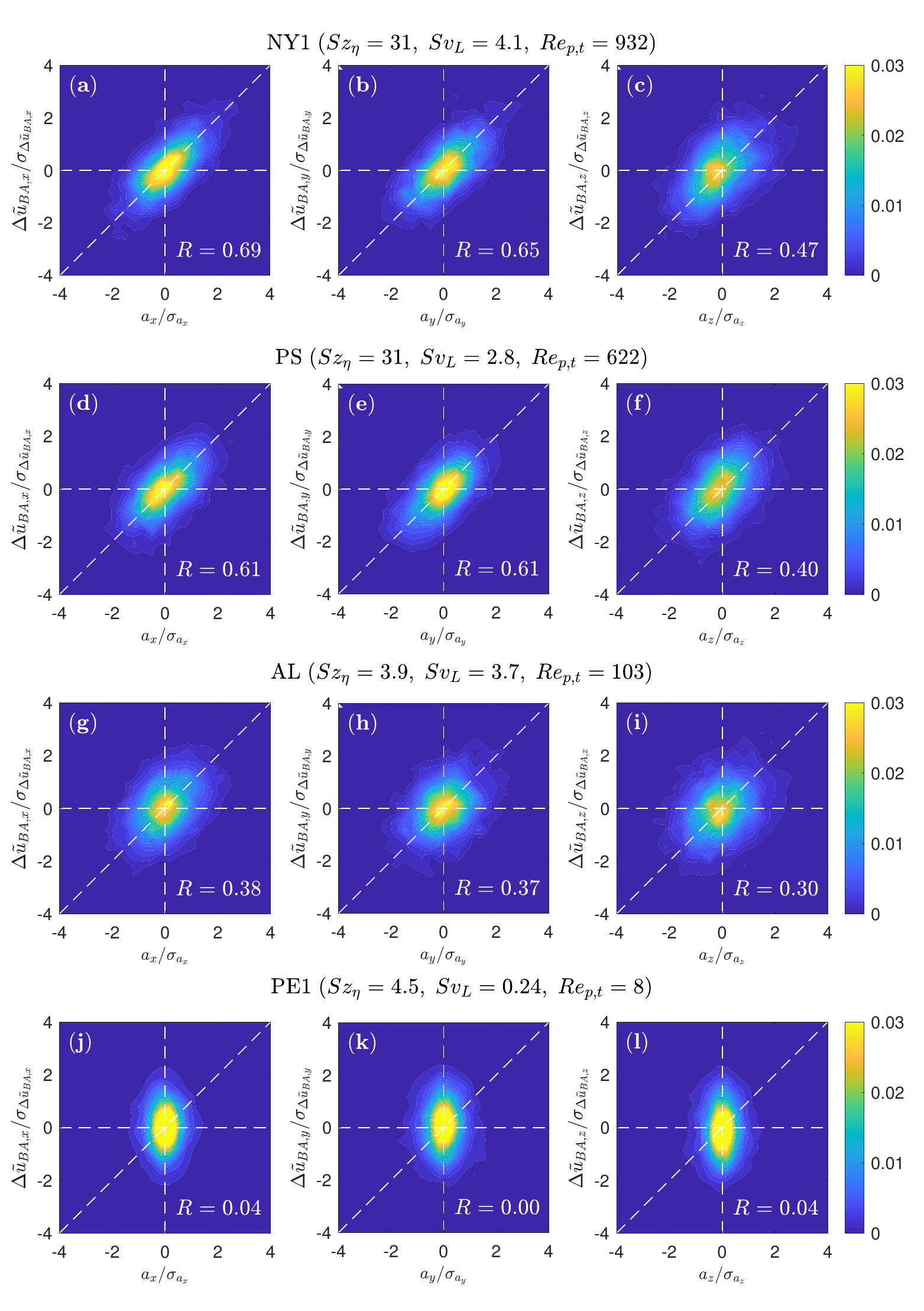}
    \caption{Joint PDFs of the instantaneous below-to-above fluid velocity difference, $\Delta \tilde{\bm{u}}_{BA}$, and particle acceleration, normalised by their respective standard deviations, for NY1, PS, AL, and PE1 particles in the (a,d,g,j) $x$, (b,e,h,k) $y$, and (c,f,i,l) $z$ directions. A time lag of $(3D_s/16)/v_s$, rounded to the nearest sampling interval, is applied to $\Delta \tilde{\bm{u}}_{BA}$ for NY1, PS, and AL, while zero lag is used for PE1. The white dashed diagonal lines indicate the one-to-one relation between the two variables, and $R$ denotes the Pearson correlation coefficient.}
    \label{fig:uba_vs_acc_NY1_PS_AL_PE1}
\end{figure}

Figure \ref{fig:uba_vs_acc_NY1_PS_AL_PE1} shows this relationship across all trajectories via ensemble-averaged joint PDFs of the instantaneous particle acceleration and the lag-corrected below-to-above fluid velocity difference. A time lag of $(3D_s/16)/v_s$ is used, rounded to the nearest sampling interval, where $3D_s/16$ is the distance from the particle centre to the centroid of either the above- or below-particle sampling hemisphere. For PE1, a zero time lag is used. The results below are not substantially affected by small changes to the chosen lag. For the NY1 particle (panels (a)--(c)), a positive correlation is observed between the two quantities in all three directions, with correlation coefficients of 0.69, 0.65, and 0.47 in the $x$, $y$, and $z$ directions, respectively. This relation is particularly evident for extreme positive and negative acceleration events, which tend to coincide with correspondingly large below-to-above fluid velocity differences. These events are consistent with particles entering or leaving the large-scale, predominantly unidirectional flow regions identified in the representative trajectory. The stronger relation in the horizontal directions may reflect the predominantly horizontal large-scale flows associated with the jet-array forcing.

Comparing across particles, we see the positive correlations are weaker for the PS particle which has a diameter similar to that of NY1 but settles more slowly (panels (d)--(f)). For the AL particle which has a slightly smaller settling velocity than NY1 but is substantially smaller (panels (g)--(i)), the relation weakens further and the joint PDFs are more scattered. And finally, for the PE1 particle which has $Sv_L < 1$ (panels (j)--(l)), no clear relation is observed between the two quantities. Together, these data suggest the crossing-trajectories effect on particle acceleration is only important for $Sv_L > 1$ and grows stronger as $Sv_L$ increases, which is consistent with expectations. Finally, it is worth pointing out that this analysis can misrepresent how particle size modulates the crossing-trajectories effect. Although the AL particle has smaller correlation coefficients compared to the PS particle despite having a larger settling velocity, this is likely reflecting the fact that by averaging over halves of the search volume, $\tilde{\bm{u}}_B$ and $\tilde{\bm{u}}_A$ filter out smaller-scale flow variations that may contribute to the acceleration of the smaller AL particle.

Overall, the analysis in the appendix supports how the crossing-trajectories effect \citep{Yudine59} sets the trends in the standard deviation and kurtosis of particle acceleration (\S\ref{sec:Particle acceleration}) for $Sv_L>1$ particles. The particle acceleration is caused by changes in the fluid velocity encountered along the settling trajectory, with the extreme events in the acceleration being even more tightly linked to the adjustment to changing flow encountered during settling.

\section{Representative trajectories of extreme acceleration events for similar sized particles}
\label{app: extreme_acc_trajectories}

\begin{figure}
    \centering
    \includegraphics[width=1\linewidth]{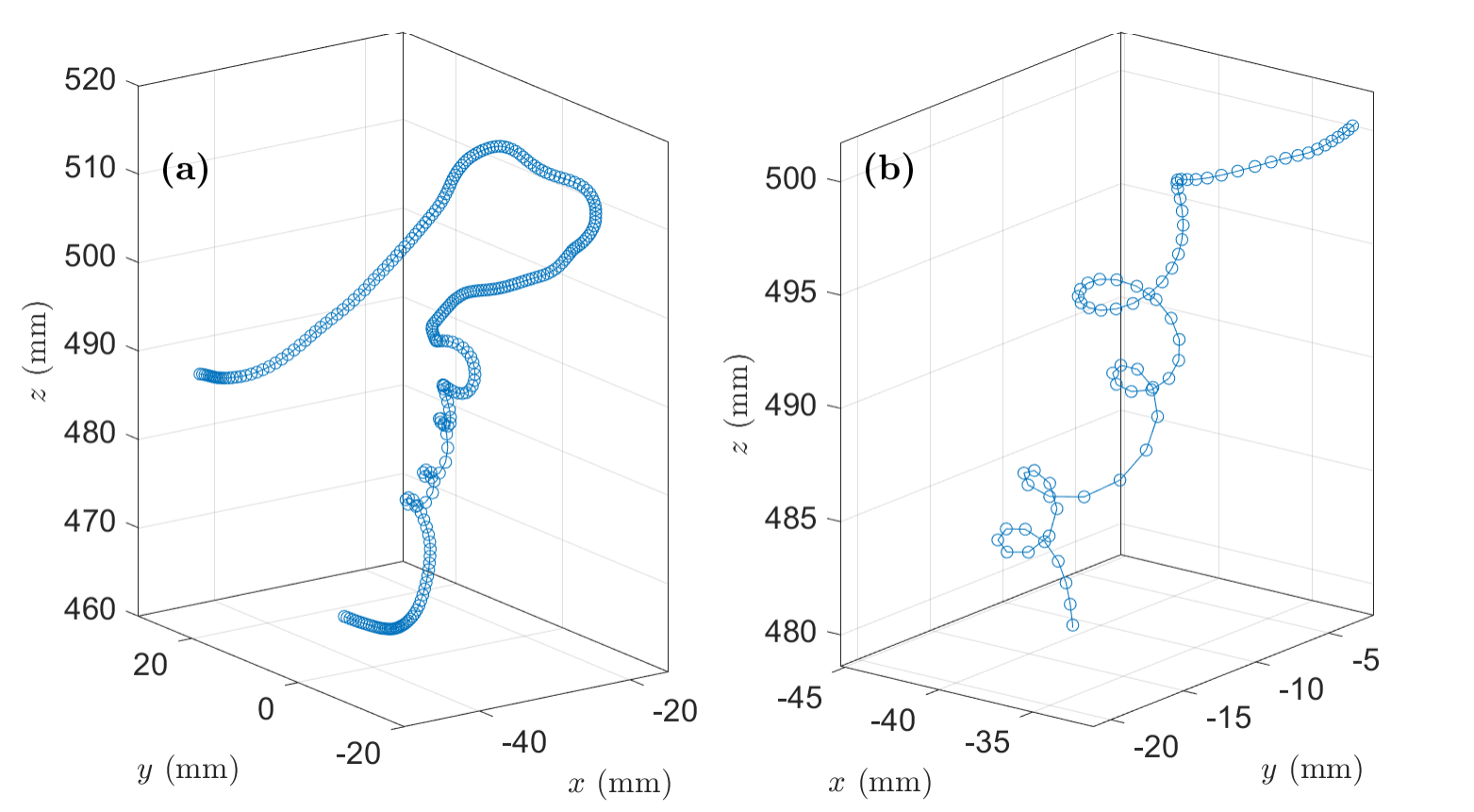}
    \caption{Representative particle trajectory for the PE1 particle with extreme particle acceleration events, where $|a_x|$ and $|a_y|$ reach approximately $0.6\textrm{g}$: (a) full trajectory and (b) zoomed-in view near the extreme acceleration events.}
    \label{fig:example_PE1_extreme_acc}
\end{figure}

\begin{figure}
    \centering
    \includegraphics[width=0.8\linewidth]{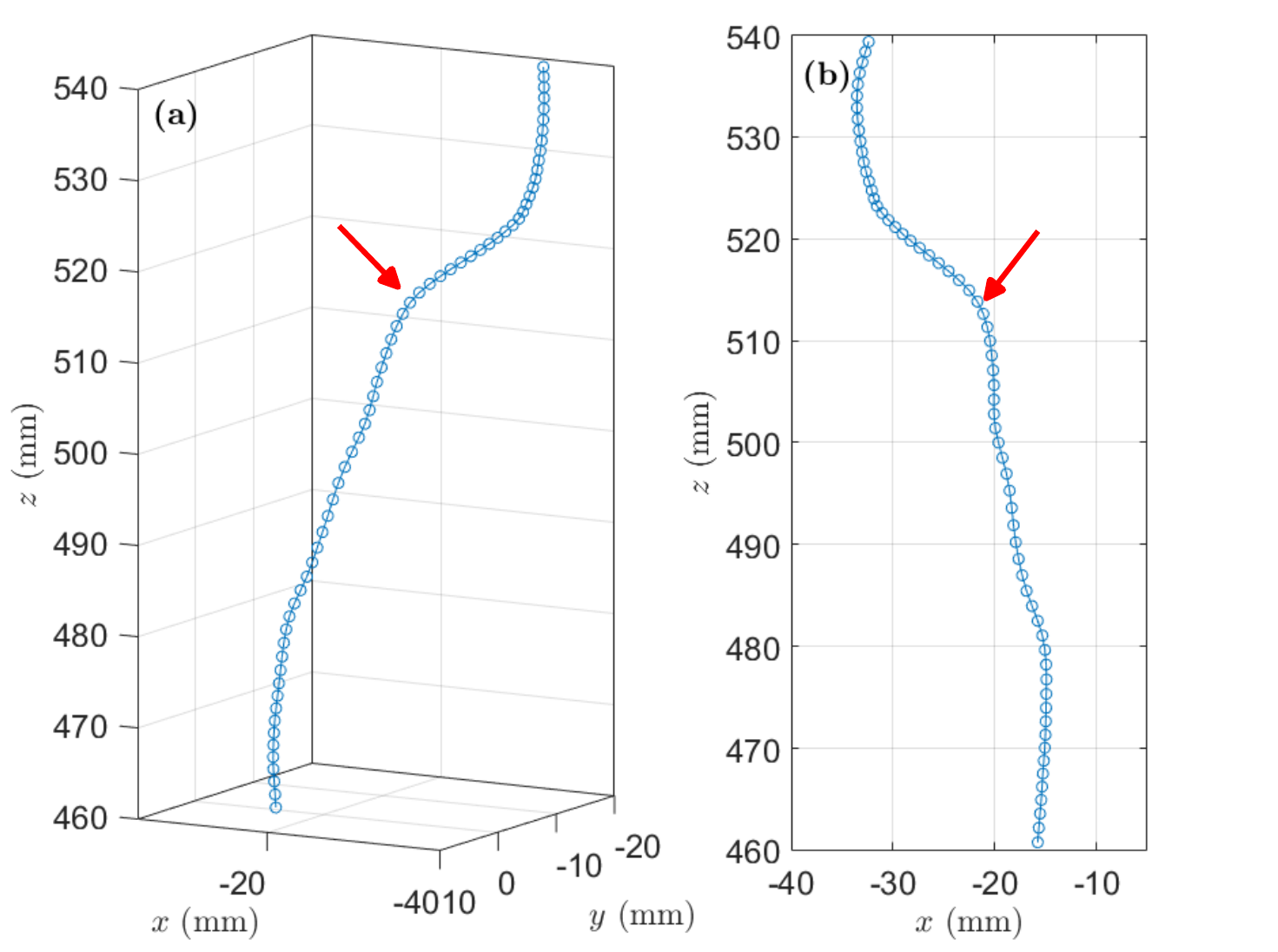}
    \caption{Representative particle trajectory of the AL particle with an extreme particle acceleration event, where $|a_x|$ reaches approximately $0.2\textrm{g}$, as indicated by the red arrows: (a) 3D view and (b) $x$-$z$ plane view.}
    \label{fig:example_AL_extreme_acc}
\end{figure}

Figures \ref{fig:example_PE1_extreme_acc} and \ref{fig:example_AL_extreme_acc} show representative trajectories of the PE1 and AL particles during extreme acceleration events. These particles are similarly sized ($Sz_\eta \approx 4.5$), but have substantially different terminal settling velocities ($Sv_L = 0.24$ \textit{vs.} $Sv_L = 3.7$). Figure \ref{fig:example_PE1_extreme_acc} shows the PE1 particle ($Sv_L = 0.24$) exhibit repeated helical motions while being advected upward. The diameter of the helical trajectory is comparable to the particle diameter, suggesting that these events could be associated with interactions between the particle and particle-scale vortical structures embedded within the large-scale upward flow. Also, this motion qualitatively resembles that reported for neutrally buoyant particles in turbulence \citep{voth2002measurement}. During the helical motion, the horizontal acceleration components reach approximately $0.6\textrm{g}$, about 20 times their respective standard deviations. In contrast, figure \ref{fig:example_AL_extreme_acc} shows the AL particle ($Sv_L = 3.7$) exhibit a change in its settling behaviour near the extreme acceleration event (indicated by red arrows) rather than being caught in a coherent turbulent eddy. Consequently, the acceleration magnitude reaches only approximately $0.2\textrm{g}$, which is still about four times the standard deviation of its acceleration.

\label{app:extreme_acc_traj}

\section{Particle-induced flow modulation and particle wakes}
\label{app:Particle_induced_flow_modulation}

In this appendix, we examine how finite-size settling particles modify the local turbulent flow. Previous studies have shown that neutrally buoyant finite-size particles can generate measurable disturbances in the surrounding flow \citep{Klein_2012,Cisse_2013}. Here, we analyse the relative motion between settling particles and surrounding tracers to determine whether there are similar spatially organised disturbance flow signatures around the particle.

To quantify the relative motion between particles and tracers, we consider the difference between the tracer velocity and the particle velocity fluctuation for each particle-tracer pair. For a tracer located at $\bm{x}_t$ and a particle centred at $\bm{x}_p$, the separation vector is defined as $\bm{r}_{tp}=\bm{x}_t-\bm{x}_p=[r_x,r_y,r_z]$, with the corresponding unit vector $\hat{\bm{r}}_{tp}=\bm{r}_{tp}/|\bm{r}_{tp}|$. In a spherical coordinate system centred on the particle, the polar angle $\theta=\cos^{-1}(r_z/|\bm{r}_{tp}|)$ is measured from the positive $z$ direction, while the azimuthal angle $\phi$ is measured from the positive $x$ direction in the horizontal $x$--$y$ plane. The corresponding unit vectors are $\hat{\bm{\theta}}=[\cos\theta\cos\phi,\cos\theta\sin\phi,-\sin\theta]$ and $\hat{\bm{\phi}}=[-\sin\phi,\cos\phi,0]$, respectively.

The tracer-particle relative velocity is then computed as $\Delta \bm{u}^{tp}=\bm{u}_t-\bm{v}^\prime$, where $\bm{u}_t$ is the instantaneous tracer velocity and $\bm{v}^\prime=\bm{v}-\langle\bm{v}\rangle$ is the particle velocity fluctuation obtained by subtracting the global mean to remove the mean settling motion. Using the spherical basis vectors defined above, the relative velocity is decomposed into radial, polar, and azimuthal components as $\Delta u_r^{tp}=\Delta \bm{u}^{tp}\cdot\hat{\bm{r}}_{tp}$, $\Delta u_\theta^{tp}=\Delta \bm{u}^{tp}\cdot\hat{\bm{\theta}}$, and $\Delta u_\phi^{tp}=\Delta \bm{u}^{tp}\cdot\hat{\bm{\phi}}$, respectively.

For each trajectory, the radial, polar, and azimuthal relative-velocity components are binned in the $(r_h,r_z)$ plane, where $r_h=(r_x^2+r_y^2)^{1/2}$ and $r_z$ are the horizontal and vertical separations from the particle centre, respectively. Following \citet{Klein_2012} and \citet{Cisse_2013}, axisymmetry about the vertical direction is assumed, and samples with the same $r_h$ and $r_z$ are averaged over the azimuthal direction, such that each bin represents an annular region around the vertical axis. The resulting trajectory-wise fields are then averaged across trajectories with equal weighting. A bin size of 1.5 mm $\times$ 1.5 mm is used to retain sufficient samples in each bin, as smaller bins produce empty regions, particularly near the vertical axis where the annular sampling area is reduced. 

\begin{figure}
    \centering
    \includegraphics[width=0.9\linewidth]{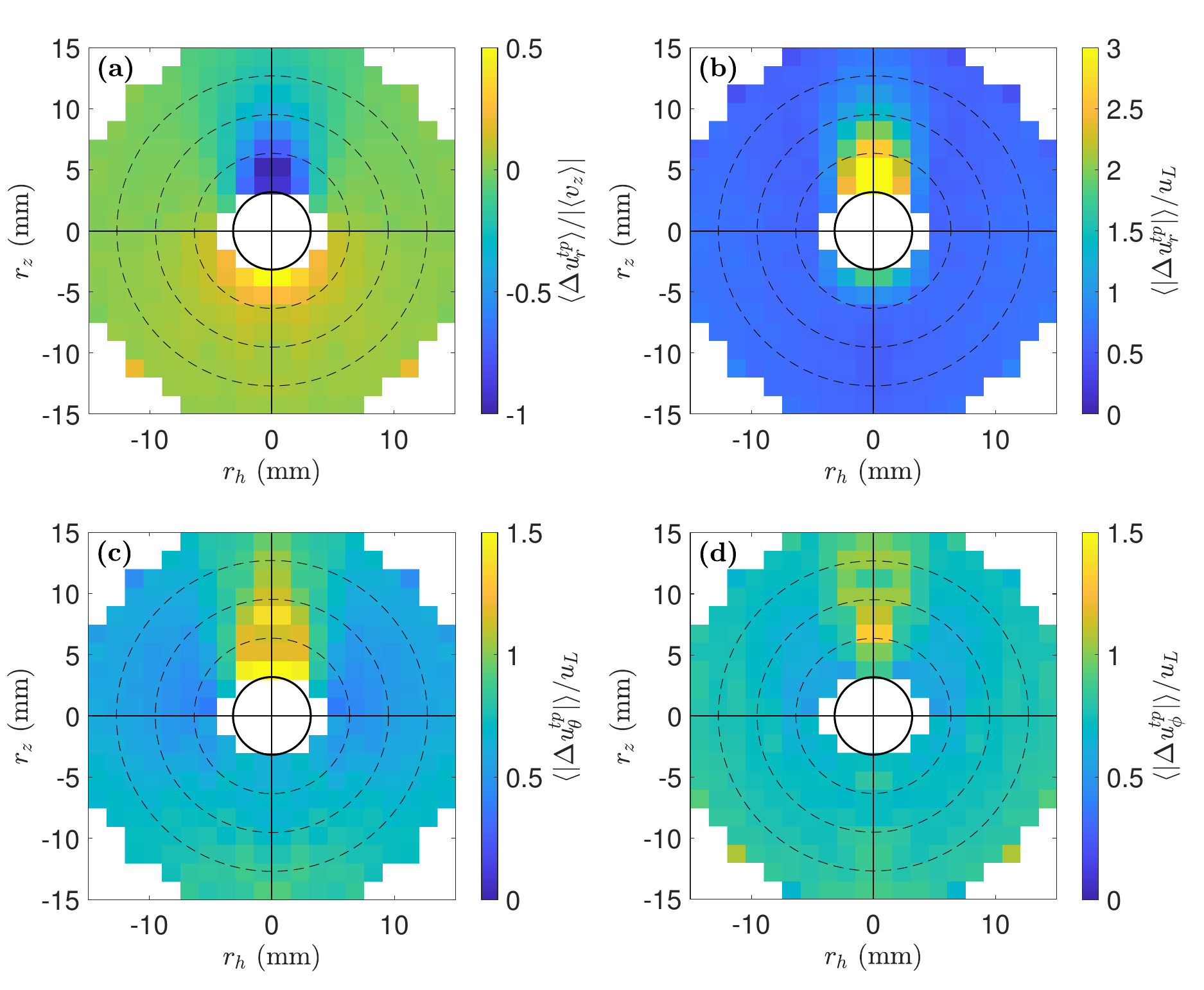}
    \caption{Particle-tracer relative motion around the NY1 particle settling in turbulence. (a) Normalised mean radial relative velocity. (b) Normalised mean absolute radial relative velocity. (c) Normalised mean absolute polar relative velocity. (d) Normalised mean absolute azimuthal relative velocity. The field in (a) is normalised by $|\langle v_z\rangle|$, whereas the fields in (b)--(d) are normalised by the large-scale turbulent velocity scale, $u_L$. The fields are computed using a bin size of 1.5 mm to ensure a sufficient number of samples in each bin. The fields for $r_h<0$ are mirrored from those for $r_h>0$. The solid black circle indicates the particle, and the dashed black circles indicate radial distances of $d_p$, $1.5d_p$, and $2d_p$ from the particle centre. The white regions indicate locations where no tracers are measured, either due to the presence of the particle or because the region lies outside the search radius, $D_s/2=15$ mm.}
    \label{fig:rel_velo_NY1}
\end{figure}

Figure \ref{fig:rel_velo_NY1} shows the particle-tracer relative motion around the NY1 particle settling in turbulence, which has the largest $d_p$ and $Re_{p,t}$. In panel (a), the mean radial relative velocity normalised by the magnitude of the global-mean vertical particle velocity, $\langle\Delta u_r^{tp}\rangle/|\langle v_z\rangle|$, exhibits distinct spatial signatures directly above and below the particle, with much smaller values elsewhere. Above the particle, $\hat{\bm{r}}_{tp}\approx \bm{e}_z$, giving $\langle\Delta u_r^{tp}\rangle \approx \langle u_{t,z}\rangle$. The negative value, with a magnitude comparable to $|\langle v_z\rangle|$, indicates a mean downward tracer velocity comparable to the mean particle settling velocity, suggesting a particle-following region. This signature weakens with increasing $r_z$ but remains observable up to at least $r_z=2d_p$ for $r_h<d_p/2$. Below the particle, $\hat{\bm{r}}_{tp}\approx -\bm{e}_z$, giving $\langle\Delta u_r^{tp}\rangle \approx -\langle u_{t,z} \rangle$. The positive value indicates a mean downward tracer velocity, suggesting that the particle displaces the fluid downward beneath it. This signature is weaker and more localised, appearing mainly within $d_p/2<|\bm{r}_{tp}|<d_p$.

Away from the regions directly above and below the particle, $\langle\Delta u_r^{tp}\rangle$ remains small. However, this may partly result from cancellation of radial relative velocities with opposite signs at different azimuthal locations during axisymmetric averaging. To avoid this cancellation, panel (b) shows the normalised mean absolute radial relative velocity, $\langle|\Delta u_r^{tp}|\rangle/u_L$. The normalised value is higher immediately above and below the particle, while elsewhere it is approximately 0.6--0.7.

Panels (c) and (d) show the normalised mean absolute polar and azimuthal relative velocities. Outside the region directly above the particle, both components are approximately 0.6–0.7, comparable to the radial component away from the particle-following region. Directly above the particle, however, both components are enhanced, with a stronger increase in the polar component. The enhanced polar component indicates substantial lateral motion towards and away from the vertical settling axis, while the azimuthal component indicates motion around this axis. The azimuthal motion could be a signature of a wake with broken axisymmetry, in which fluid motion is no longer confined to the radial and axial directions \citep{GHIDERSA_2000}. These observations suggest that the non-axisymmetric wake dynamics associated with path instability, previously observed in quiescent fluid, persist in turbulence.

\begin{figure}
    \centering
    \includegraphics[width=0.9\linewidth]{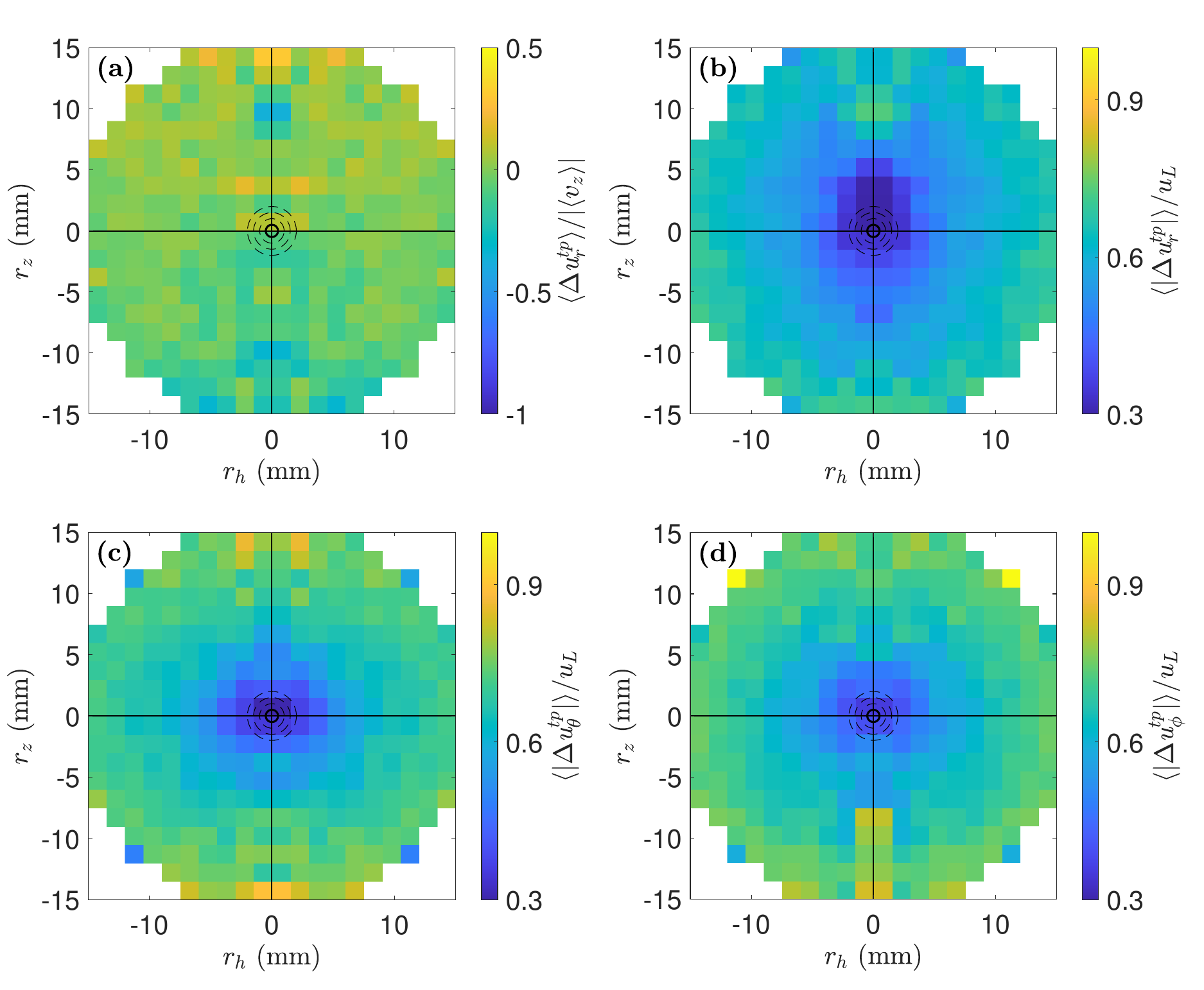}
    \caption{Particle-tracer relative motion around the PE2 particle settling in turbulence. Description is the same as figure \ref{fig:rel_velo_NY1}. }
    \label{fig:rel_velo_PE2}
\end{figure}

In contrast, figure \ref{fig:rel_velo_PE2} shows the particle-tracer relative motion around the PE2 particle, which has the second-smallest settling velocity among the particles considered but still has a sufficiently large mean settling velocity for normalisation. Because the bin size exceeds the particle diameter, flow structures very close to the particle are not fully resolved. Nevertheless, the mean absolute radial, polar, and azimuthal relative velocities all decrease towards the particle, indicating smaller differences between the particle velocity fluctuations and the surrounding tracer velocities and suggesting stronger coupling between the particle and nearby fluid motion.

Direct assessment of flow modification and detailed characterisation of near-particle flows are limited by the present spatial resolution. However, similar spatial structures, characterised by fore-aft asymmetry and an extended disturbance behind the particles, are observed around the two largest particles, NY1 and PS, consistent with local modifications of the surrounding turbulent flow by finite-size settling particles. Because both particles have $Re_{p,t} > 210$, above the path-instability threshold, these observations further suggest that wake-like flow structures associated with path instability in quiescent fluid may persist in turbulence.

Finally, we assess how the estimation of $\tilde{\bm{u}}$ is affected by the particle-induced flow modulation. We choose the NY1 particle, which is the faster-settling of the two largest particles and exhibits the strongest flow modulation. Excluding tracers within the region $r_z > -d_p$ and $0 \leq r_h \leq d_p$ leaves the statistics of $\tilde{\bm{u}}$ largely unchanged, except for $\langle \tilde{u}_z \rangle$, which increases by approximately 7\% of $u_L$ and 2\% of $v_s$, consistent with the exclusion of the downward flow induced by the settling particle.

\section{Particle motion relative to the local coarse-grained flow}
\label{app:v_u_mean_std}

\begin{figure}
    \centering
    \includegraphics[width=\linewidth]{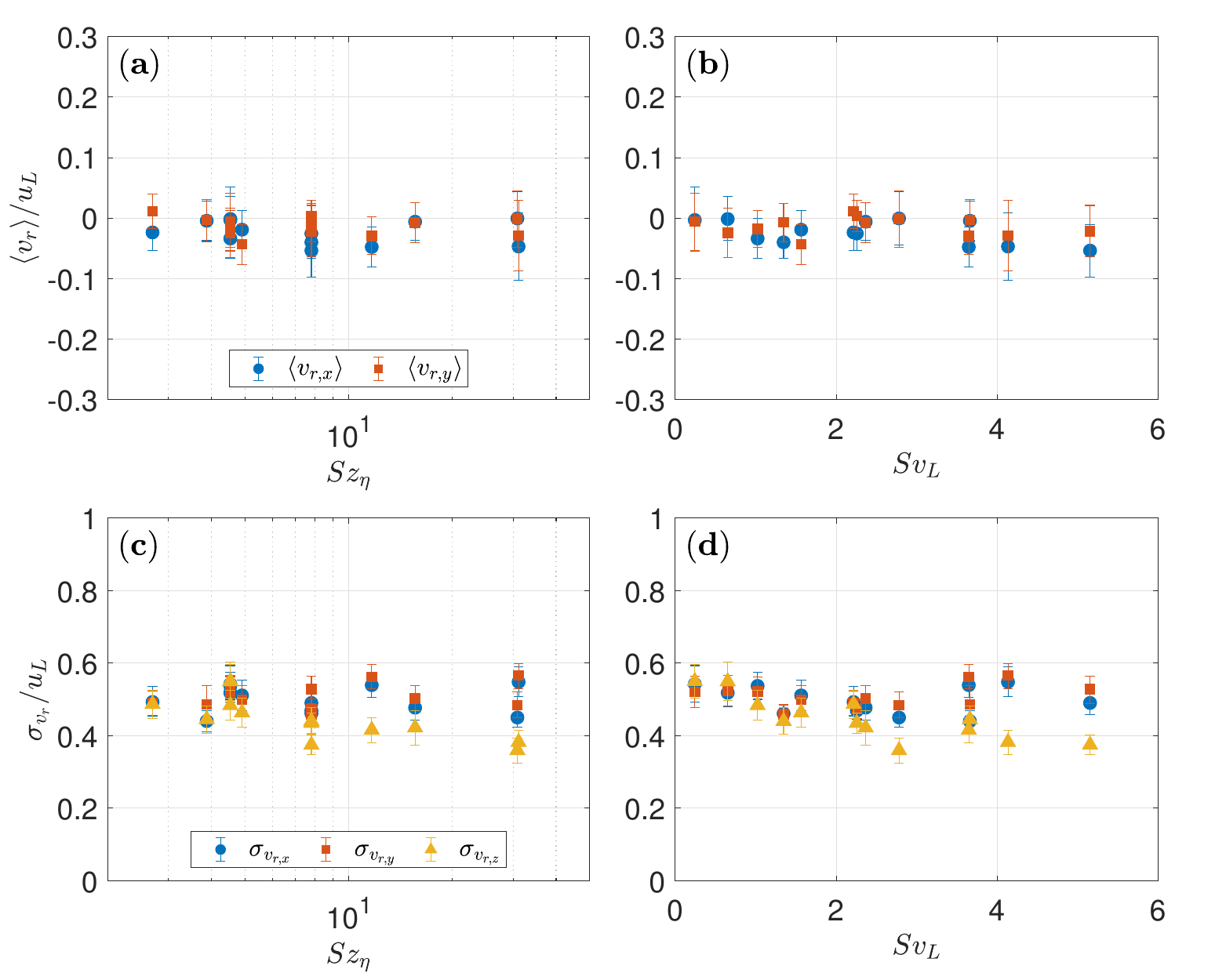}
    \caption{Normalised settling-corrected mean relative velocities as functions of (a) $Sz_\eta$ and (b) $Sv_L$, and normalised standard deviations of settling-corrected relative velocities as functions of (c) $Sz_\eta$ and (d) $Sv_L$. Error bars represent 95 \% confidence intervals obtained by bootstrapping.}
    \label{fig:mean_rms_slip_Sz_Sv}
\end{figure}

\begin{figure}
    \centering
    \includegraphics[width=0.8\linewidth]{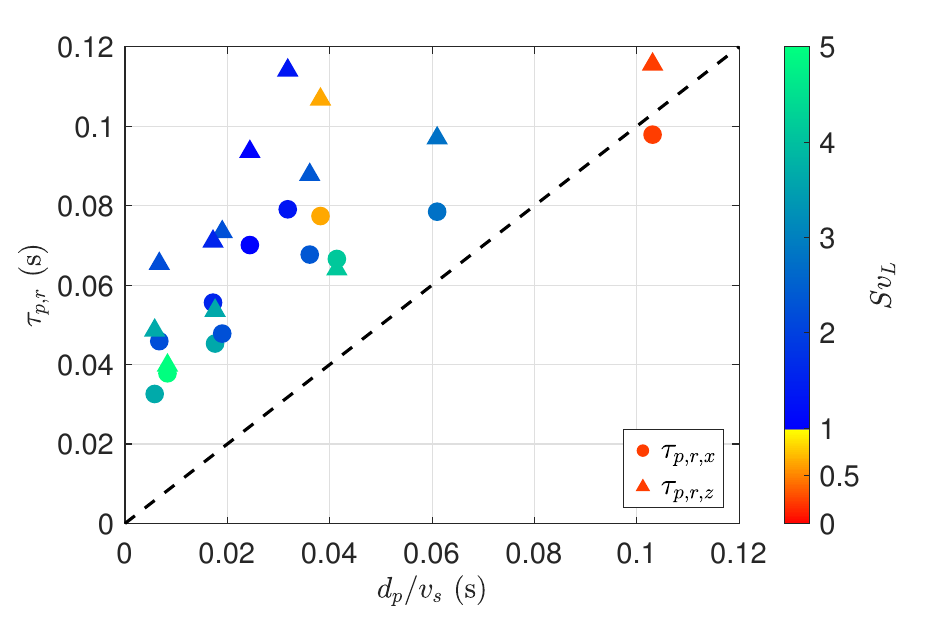}
    \caption{Relative velocity timescale as a function of $d_p/v_s$. Circle and triangle markers represent the $x$- and $z$-directional components, respectively. Marker colour indicates $Sv_L$. For visual clarity, separate colormaps are used for particles with $Sv_L<1$ and $Sv_L>1$. The black dashed line indicates the one-to-one relation.}
    \label{fig:tau_p_r_vs_est.eps}
\end{figure}

In this appendix, we analyse the statistics of the settling-corrected relative velocity with respect to the local coarse-grained flow, $\bm{v}_r \equiv \bm{v} - \tilde{\bm{u}} + v_s \bm{e}_z$. We focus the analysis on the relative flow between the particle and the local fluid that is distinct from settling (including how settling is modified by turbulence). This analysis therefore quantifies the integrated effects of how the particle responds to the forcing it receives from the local fluid motion. 

Figure \ref{fig:mean_rms_slip_Sz_Sv} shows the mean and standard deviations of $\bm{v}_r$ as functions of $Sz_\eta$ and $Sv_L$, with only the horizontal components shown for the mean. (The vertical component of the mean is related to the settling modulation and discussed in \S\ref{sec:Particle_velocity}.) The data in panels (a) and (b) clearly show that the mean horizontal particle motion reflects the mean fluid motion coarse-grained at the integral scale. In panels (c) and (d), the standard deviations of $\bm{v}_r$ are approximately $0.5 u_L$ across all particles, with no clear dependence on $Sz_\eta$ and $Sv_L$. This suggests that the coarse-grained slip is set by how particles adjust to the integral-scale turbulent motions without any specific role played by the particle size or settling velocity. The vertical component exhibits a modest decrease with increasing $Sv_L$, showing that faster settling particles experience smaller fluctuations in their settling-corrected relative velocity. However, this decrease is primarily due to a decrease in the particle velocity fluctuations and not related to fluctuations in the sampled fluid velocity (not shown).

Figure \ref{fig:tau_p_r_vs_est.eps} shows the timescale of the relative-velocity autocorrelation $\tau_{p,r}$ (computed as described in Appendix \ref{app:autocorrelation}). This timescale therefore quantifies the delay over which the particle reaches equilibrium with the settling-corrected local flow. Its value is larger than the particle acceleration timescale $\tau_{p,a}$ (figure \ref{fig:tau_p_a_vs_tau_p_est}) since the particle velocity is the integral of the particle acceleration. We observe that $\tau_{p,r}$ is substantially smaller than the integral timescale of the turbulence, $T=1.03$ s for all particles, indicating that the particle comes to equilibrium with the local integral-scale fluid motion substantially faster than the characteristic evolution of that flow.

Together, these analyses suggest that the particles in our dataset do not exhibit an inertial response with respect to the integral scale turbulent eddies.

\section{Drag analysis of settling particles based on \citet{Fornari_2016_reduced}}
\label{app:Fornari_drag_analysis}

\citet{Fornari_2016_reduced} introduced a method to analyse the hydrodynamic force $\bm{F}_{\textrm{hd}}$ in Eq.~\eqref{eq:particle_force}. Under the simplification that mean particle acceleration vanishes (supported by the data; see figure \ref{fig:mean_acc_Sz_Sv}), the mean vertical hydrodynamic force is balanced by the submerged weight of the particle, $\langle {F}_{\textrm{hd},z} \rangle = (\gamma - 1)\rho\mathrm{g}(\pi d_p^3/6 )$. This mean vertical hydrodynamic force is then modelled as the sum of `steady' and `unsteady' terms, where the `steady' term is written using standard drag correlations and the `unsteady' term makes up the difference. 

We follow this analysis using the Schiller--Naumann drag correlation for the `steady' component, evaluated using the instantaneous particle--fluid relative velocity under the quasi-steady assumption. Then the dimensionless budget of the mean vertical hydrodynamic force $\langle {F}_{\textrm{hd},z} \rangle$ can be written as \citep{Fornari_2016_reduced}
\begin{equation}
    1 = \underbrace{\frac{(v_s-\langle v_{r,z}\rangle)(1+\hat{K})}{V_S}}_{\text{Mean settling}}  +\underbrace{\frac{(v_s-\langle v_{r,z}\rangle)K''}{V_S}}_{\text{Cross-flow}} - \underbrace{\frac{\langle v'_{r,z}K'\rangle} {V_S}}_{\text{Nonlinear}}     +\underbrace{f_D^U}_{\text{Unsteady}},
    \label{eq:Fornari_drag_budget}
\end{equation}
where $\bm{v}_r$ is defined in Eq.~\eqref{eq:kinematic_model_particle_velo}, $V_S=(\gamma-1)gd_p^2/(18\nu)$ is the Stokes settling velocity, and 
\begin{equation}
    K=0.15\left(\frac{d_p|\bm{v}_r-v_s\bm{e}_z|}{\nu}\right)^{0.687},
   \end{equation}
is the `correction factor' from Stokes drag in the Schiller--Naumann drag correlation. This correction factor is then decomposed as $K = \hat{K} + K' + K''$, where $K'=K-\langle K\rangle$ represents the fluctuation of the drag correction; $K''=\langle K\rangle-\hat{K}$ represents the difference between the mean drag correction and the drag correction evaluated using the mean vertical relative velocity, with $\hat{K}=0.15( d_p|\langle v_{r,z}\rangle-v_s|/{\nu})^{0.687}$. Note that the negative sign in front of the nonlinear drag contribution arises from our choice of an upward-positive vertical coordinate, in contrast to the downward-positive convention adopted by \citet{Fornari_2016_reduced}. Thus, in Eq.~\eqref{eq:Fornari_drag_budget}, the first three terms on the right-hand side are derived from the `steady' component and can be evaluated from the data, whereas the unsteady component $f_D^U$ is assumed to be the residual.

\begin{figure}
    \centering
    \includegraphics[width=\linewidth]{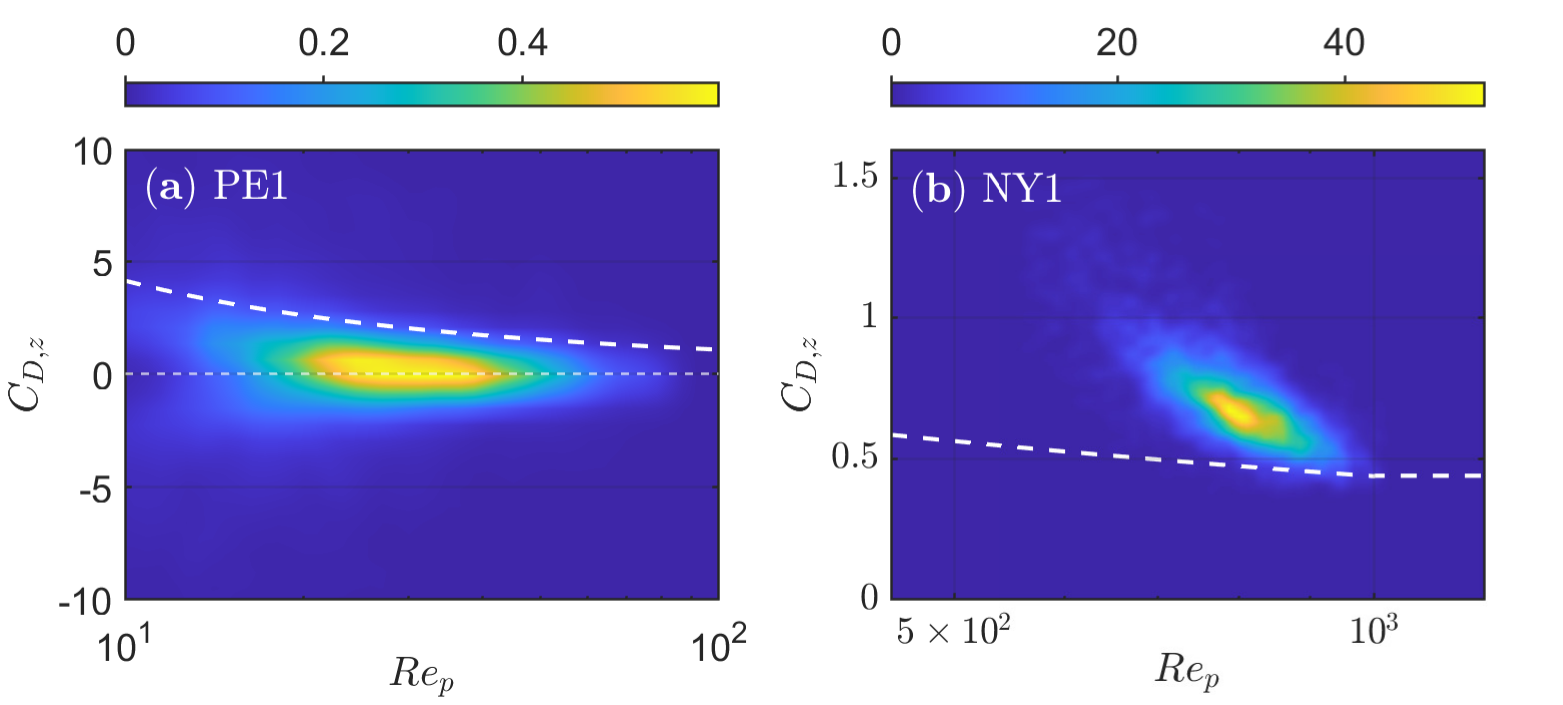}
    \caption{Joint probability density functions of the instantaneous vertical drag coefficient $C_{D,z}$ and the instantaneous particle Reynolds number $Re_p = |\bm{v}_r-v_s \bm{e}_z|d_p/\nu$ for (a) PE1 and (b) NY1. The white dashed curves represent the Schiller--Naumann drag model, with $C_D=0.44$ for $Re_p>1000$.}
    \label{fig:drag_coeff_PE1_NY1}
\end{figure}

\begin{figure}
    \centering
    \includegraphics[width=0.7\linewidth]{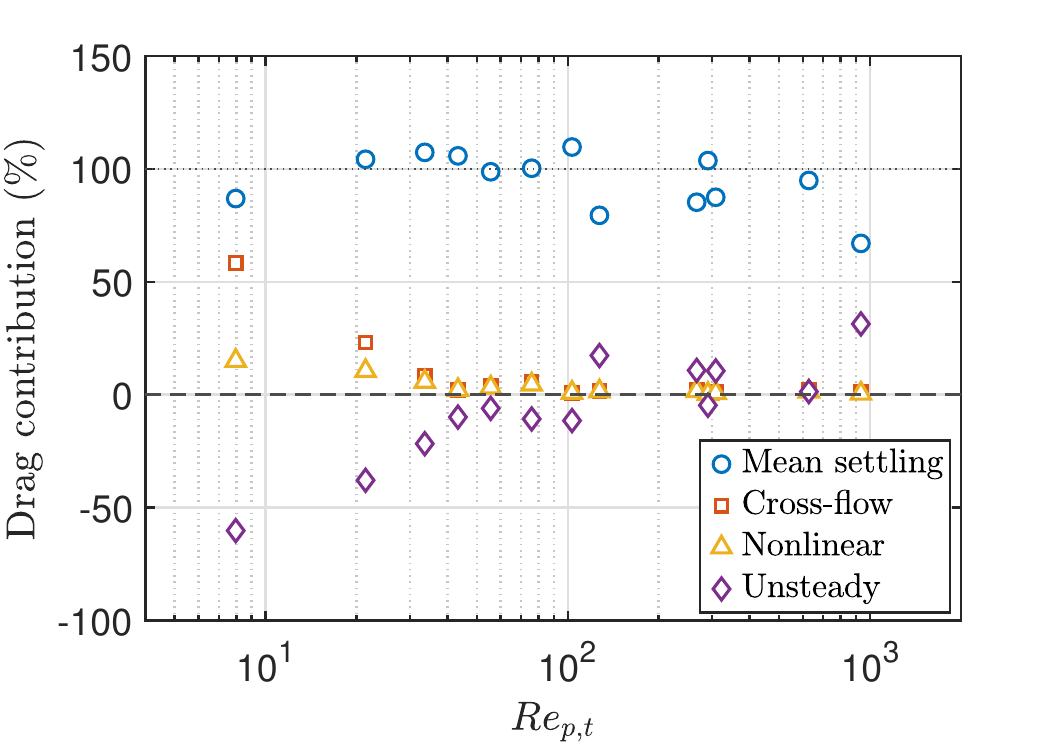}
    \caption{Contributions of the mean-settling, cross-flow, nonlinear and unsteady drag terms, normalised by the submerged weight of the particle, as a function of the terminal particle Reynolds number $Re_{p,t}$. The drag contributions are estimated using the Schiller--Naumann drag model. The horizontal dashed and dotted lines indicate 0\% and 100\%, respectively.}
    \label{fig:drag_contribution}
\end{figure}

Before examining the data with this analysis method, we first use the data to examine how the instantaneous vertical drag force on the particle relates to the Schiller--Naumann drag correlation. Figure~\ref{fig:drag_coeff_PE1_NY1} plots the distribution of the instantaneous vertical drag coefficient $C_{D,z}=-F_{\textrm{hd},z}/[\tfrac{1}{2}\rho A_p|\bm{v}_r-v_s\bm{e}_z|(v_{r,z}-v_s)]$ as a function of the instantaneous particle Reynolds number $Re_p=d_p|\bm{v}_r-v_s\bm{e}_z|/\nu$ for PE1 and NY1 particles. Here, $A_p=\pi d_p^2/4$ is the projected area of the particle and $F_{hd,z}$ is the vertical component of the hydrodynamic force. We observe that the Schiller--Naumann drag correlation overestimates the instantaneous vertical drag for the PE1 particle whereas it underestimates it for the NY1 particle. In both cases, the data suggest that the drag coefficient for settling in turbulence is qualitatively a different function of the instantaneous particle Reynolds number than standard drag correlations found in steady conditions with a uniform free stream flow incident onto a fixed particle. Therefore, a quasi-steady assumption is unlikely to be useful, and in the case of the PE1 particle, likely to suggest that the `unsteady' component is negative. 

Figure \ref{fig:drag_contribution} shows the drag budget for all particles as a function of $Re_{p,t}$ based on Eq.~\eqref{eq:Fornari_drag_budget}. Consistent with the PE1 data in figure \ref{fig:drag_coeff_PE1_NY1}, we find negative values for the unsteady component for $Re_{p,t} \lesssim 100$. Despite the significant contributions of the cross-flow and nonlinear terms for $Re_{p,t} \lesssim 100$ particles, our conclusion is that the quasi-steady assumption is not a useful model for this analysis. For $Re_{p,t} \gtrsim 100$ particles, the unsteady component is near-zero or positive, but the nonlinear and cross-flow terms are negligible \citep[consistent with previous findings;][]{Fornari_2016_reduced,Clementi_2026}. Thus, the analysis does not reveal useful insights for this regime either, and the settling modulation due to turbulence in this regime is better understood in terms of changes to the particle wake (\S\ref{sec:Particle_velocity}). 


\backsection[Acknowledgements]{The authors gratefully acknowledge Alejandro Roldán Alzate and James Rice at UW-Madison for use of their high-speed camera and Greg Voth at Wesleyan University for stimulating discussions.}

\backsection[Funding]{This research was supported by the US National Science Foundation (ENG-2211704 and ENG-2338221) awarded to N.P. and a FACE Foundation / Transatlantic Research Partnership grant awarded to N.P. and G. V. Additionally, N.P. acknowledges an Early-Career Research Fellowship from the Gulf Research Program of the National Academies of Science, Engineering, and Medicine.}

\backsection[Declaration of interests]{The authors report no conflict of interest.}

\backsection[Data availability statement]{The data that support the findings of this study are openly available in [repository name] at http://doi.org/[doi], reference number [reference number]. [To be filled upon acceptance]}

\backsection[Author ORCIDs]{Joo Young Bang, https://orcid.org/0009-0000-3302-5659; Gautier Verhille, https://orcid.org/0000-0001-6521-6433; Nimish Pujara, https://orcid.org/0000-0002-0274-4527}

\backsection[Author contributions]{N.P. and J.Y.B. designed the study, J.Y.B. conducted the experiments and performed data analysis with help from G.V. and N.P., all authors contributed to the interpretation of results and the writing of the manuscript.}

\bibliographystyle{jfm}
\bibliography{reference}

\end{document}